\documentstyle[aps,multicol,tighten,eqsecnum,psfig]{revtex}%
\begin{document}
\draft
\title{Atom Lasers, Coherent States, and Coherence: \\
II. Maximally Robust Ensembles of Pure States}
\author{H.M. Wiseman$^{1,3,2,4,}$
\footnote{Electronic address: h.wiseman@gu.edu.au} and John A.
Vaccaro$^{2,4,1}$}
\address{$^{1}$School of Science, Griffith University, Brisbane 4111
Australia.}
\address{$^{2}$Division of Physics and Astronomy, University of
Hertfordshire, Hatfield AL10 9AB, UK.}
\address{$^{3}$Department of Physics, University of Queensland, Queensland
4072 Australia.}
\address{$^{4}$Physics Department, The Open University, Milton Keynes
MK7 6AA,  United Kingdom}
\maketitle

\begin{abstract}
As discussed in the preceding paper [Wiseman and Vaccaro, quant-ph/9906125],
the stationary state of an optical or atom laser far above
threshold is a mixture of coherent field states with random phase,
or, equivalently, a Poissonian mixture of number states. We are
interested in which, if either, of these descriptions of
$\rho_{\rm ss}$ as a stationary ensemble of pure states, is more
natural. In the preceding paper we concentrated upon the question
of whether descriptions such as these are physically realizable
(PR). In this paper we investigate another relevant aspect of
these ensembles, their robustness. A robust ensemble is one for
which the pure states that comprise it survive relatively
unchanged for a long time under the system evolution. We determine
numerically the most robust ensembles as a function of the
parameters in the laser model: the self-energy $\chi$ of the
bosons in the laser mode, and the excess phase noise $\nu$. We
find that these most robust ensembles are PR ensembles, or similar
to PR ensembles, for all values of these parameters. In the ideal
laser limit ($\nu=\chi=0$), the most robust states are coherent
states. As the phase noise or phase dispersion is increased
through $\nu$ or the self-interaction of the bosons $\chi$,
respectively, the most robust states become more and more
amplitude-squeezed. We find scaling laws for these states, and
give analytical derivations for them. As the phase diffusion or
dispersion becomes so large that the laser output is no longer
quantum coherent, the most robust states become so squeezed that
they cease to have a well-defined coherent amplitude. That is, the
quantum coherence of the laser output is manifest in the most
robust PR ensemble being an ensemble of states with a well-defined
coherent amplitude. This lends support to our approach of
regarding robust PR ensembles as the most natural description of
the state of the laser mode.  It also has interesting implications
for atom lasers in particular, for which phase dispersion due to
self-interactions is expected to be large.

\end{abstract}


\newcommand{\beq}{\begin{equation}}
\newcommand{\eeq}{\end{equation}}
\newcommand{\bqa}{\begin{eqnarray}}
\newcommand{\eqa}{\end{eqnarray}}
\newcommand{\nn}{\nonumber}
\newcommand{\nl}{\nn \\ &&}
\newcommand{\dg}{^\dagger}
\newcommand{\erf}[1]{Eq.~(\ref{#1})}
\newcommand{\smallfrac}[2]{\mbox{$\frac{#1}{#2}$}}
\newcommand{\bra}[1]{\left\langle{#1}\right|}
\newcommand{\ket}[1]{\left|{#1}\right\rangle}
\newcommand{\ip}[1]{\left\langle{#1}\right\rangle}
\newcommand{\sch}{Schr\"odinger }
\newcommand{\schs}{Schr\"odinger's }
\newcommand{\hei}{Heisenberg }
\newcommand{\heis}{Heisenberg's }
\newcommand{\half}{\smallfrac{1}{2}}
\newcommand{\bl}{{\bigl(}}
\newcommand{\br}{{\bigr)}}
\newcommand{\ito}{It\^o }
\newcommand{\tr}[1]{{\rm Tr}\left[ {#1} \right]}
\newcommand{\singlecol}{\end{multicols}
     \vspace{-0.5cm}\noindent\rule{0.5\textwidth}{0.4pt}\rule{0.4pt}
     {\baselineskip}\widetext }
\newcommand{\doublecol}{\noindent\hspace{0.5\textwidth}
     \rule{0.4pt}{\baselineskip}\rule[\baselineskip]
     {0.5\textwidth}{0.4pt}\vspace{-0.5cm}\begin{multicols}{2}\noindent}

\begin{multicols}{2}
\narrowtext

\section{Introduction}

A laser is a device that produces a coherent beam of bosons. The
meaning of the word `coherent' in this context is discussed at
length in a paper by one of us \cite{Wis97}. In particular, a
coherent output does not mean that the output, or the laser mode
itself, is in a coherent state. Rather, as has long been
recognized \cite{SarScuLam74}, the stationary state matrix for the
laser mode is a mixture of number states. In the far-above
threshold limit, this mixture is Poissonian with mean $\mu$:
\beq
   \rho_{\rm ss} = \sum_{n=0}^{\infty} e^{-\mu}\frac{\mu^{n}}{n!}\ket{n}\bra{n} .
   \label{mixnum} \label{rhoinf}\eeq
This state matrix can also be represented as a mixture of coherent
states:
\beq
  \rho_{\rm ss} = \int \frac{d\phi}{2\pi}
  \ket{|\alpha|e^{i\phi}}\bra{|\alpha|e^{i\phi}}, \label{mixcoh}
\eeq
where $|\alpha|^{2}=\mu$.

On the basis of this second representation, one might claim that
the laser really is in a coherent state $\ket{|\alpha|e^{i\phi}}$,
but that one cannot know {\em a priori} what the phase $\phi$ is.
In the preceding paper \cite{WisVac01a} we have investigated
whether this claim is true. If it were true then there should be
some way of finding out which coherent state the laser is in
without affecting its dynamics. We found that if there is any
self-energy  in the laser mode (such as a $\chi^{(3)}$
nonlinearity for an optical laser, or $s$-wave scattering for an
atom laser), then it is in fact {\em not} possible to physically
realize the coherent state ensemble in \erf{mixcoh}. By contrast,
it is always possible to physically realize the number state
ensemble in \erf{mixnum}.

For an ideal laser (with no $\chi^{(3)}$-like nonlinearity), the
unknown coherent state description and the unknown number state
description are both physically realizable (PR). Given that they
are mathematically equivalent, why is the former description
ubiquitous and the latter rare? The answer, as was pointed out
some time ago by Gea-Banacloche \cite{Gea90}, is differential
survival times. An ideal laser prepared in a coherent state will
remain close to that initial state for a time of order
$\kappa^{-1}$, where $\kappa$ is the bare decay rate of the
cavity. By contrast, a laser prepared in a number state will be
likely to remain in that state only for a time of order
$\kappa^{-1}/\mu$, where $\mu$ is the mean number as above.

This result, derived also in Ref.~\cite{Wis93a}, was taken further
by Gea-Banacloche in Ref.~\cite{Gea98} using the early model for a
laser with saturation due to Scully and Lamb \cite{SarScuLam74}.
Gea-Banacloche considered pure states with mean photon number
equal to that of the laser at steady state, and calculated their
purity at later times. He showed that the pure state that had the
slowest initial rate of decay of purity was, in general, a
slightly amplitude-squeezed state rather than a coherent state.

There seems little doubt, then, that it is most useful to consider
an ideal laser to be in a coherent state (or nearly coherent
state) of unknown phase. However it is an open question whether
this is true of a non-ideal laser, that is, a laser with
additional noise or dispersion of some form. Another open question
is how this issue relates to the quantum coherence of the output
of such a non-ideal device.

The particular laser system of interest here is the atom laser
\cite{Wis97}. An important difference between an atom laser and an
optical laser is that the interatomic interactions cannot be
neglected. This gives rise to a $\chi^{(3)}$-like nonlinearity in
the laser mode. As noted above, this affects the physical
realizability of ensembles, and we also expect it to affect their
robustness.

A robustness analysis for a Bose-Einstein condensate has been done
by one of us with Barnett and Burnett \cite{BarBurVac96}. This
produced similar results to that of Gea-Banacloche \cite{Gea98},
although it was based on the fidelity \cite{Schumacher} which
measures the overlap of the initial state with the state at a
later time. However, the authors of Ref.~\cite{BarBurVac96} only
calculated the initial rate of decay of the fidelity, and this is
unaffected by any Hamiltonian terms. Hence the self energy played
no role in this analysis. Moreover, the treatment, like that of
Gea-Banacloche~\cite{Gea98}, considered only a single pure state
to represent the state of the condensate. Thus it does not give,
in general, a representation of the steady state on par with
\erf{mixnum} or \erf{mixcoh}.

In this paper we give an analysis that treats the dynamics of an
atom laser at all times and that incorporates an ensemble of pure
states.  It takes into account Hamiltonian terms and gives a
robust representation of the steady state. We consider both the
problem of finding the most robust ensemble, and the most robust
physically realizable (PR) ensemble. Since ensembles are realized
by unraveling the master equation \cite{WisVac98,WisVac01a},
finding the most robust PR ensemble is equivalent to finding the
{\em maximally robust unraveling}, a concept introduced by us in
Ref.~\cite{WisVac98}.

A review of maximally robust unravelings, including a comparison
with other approaches, is given in Sec.~II. In Sec.~III we present
the equations for determining the maximally robust unraveling
(MRU) for an atom laser model. We concentrate upon continuous
Markovian unravelings, which give ensembles of Gaussian states,
and also consider unconstrained Gaussian ensembles. In Sec.~IV we
present the numerical solutions for these equations, concentrating
on the asymptotic behaviour in the limit of large nonlinearity
$\chi$ and phase noise $\nu$. The concluding Sec.~V is a
discussion of our results and their relation to atom laser
coherence, and some suggestions for future work.

\section{Maximally Robust Unravelings}

\subsection{Comparison with Other Approaches}

The idea of robustness has it origins in studies of decoherence
and the classical limit \cite{Gea90,Wis93a,Gea98,BarBurVac96,%
Zur93,ZurHabPaz93,Gal95,ParScu98,DioKie00,DalDziZur01}.
Decoherence is the process by which an open quantum system becomes
entangled with its environment, thereby causing its state to
become mixed.  However, not all pure states decohere with equal
rapidity.  In particular, Zurek \cite{Zur93} defined the
``preferred states'' of open quantum systems as those states that
remain relatively pure for a long time. This idea can be thought
of as a ``predictability sieve'' \cite{ZurHabPaz93}. That is, the
preferred states are those for which the future dynamics are
predictable, in the sense that there is some projective question
(is the system in some particular state?) that is  likely to give
the result ``yes''.

Our approach, as introduced in Ref.~\cite{WisVac98} and applied to
resonance fluorescence by one of us and Brady \cite{WisBra00}, is
to find the maximally robust unraveling. This approach shares some
similarities with other approaches. It has, however, a suite of
four distinctive characteristics which we enumerate below.

\subsubsection{Ensembles of Pure States}

First, we considered not a single   pure state, but an ensemble of
pure states. This is appropriate for situations where the open
system comes to a mixed equilibrium state. The ensemble of pure
states that we consider must be a  representation of that
equilibrium mixed state. That is, the system has a certain
probability of  being in one of those pure states, as in
Eqs.~(\ref{mixnum}) and (\ref{mixcoh}). Recently, Di\'osi and
Kiefer \cite{DioKie00} have also considered ensembles of pure
states in a similar context.

Without considering such an ensemble it is necessary to put some
ad-hoc restriction on the pure states considered so that they have
some relevance to the actual state the system is in at
equilibrium. For example, as noted above, Gea-Banacloche
\cite{Gea98} considered  only pure states having the same mean
photon number as the equilibrium state of the laser model under
consideration.

\subsubsection{Physical Realizability} \label{secpr}

Second, in Ref.~\cite{WisVac98} we placed a restriction on the
ensembles of pure states that we consider: they  must be
physically realizable.  By this we mean that it should be
possible, without altering the evolution of the system, to know
that its state at equilibrium is definitely one of the pure states
in the ensemble, but {\em which} pure state cannot be predicted
beforehand. Di\'osi and Kiefer \cite{DioKie00} have considered a
similar condition, although they do not make the connection with
physical realizability and measurement. In this paper we also
consider ensembles without the constraint of physical
realizability, as it is of interest to see how active that
constraint is.

We have considered in detail the issue of physical realizability
of ensembles of pure states in the preceding paper
\cite{WisVac01a}. Here we merely remind the reader of some key
points and terminology. An ensemble for a system obeying a
Markovian master equation is physically realized by monitoring the
baths to which it is coupled. This leads to an {\em unraveling}
\cite{Car93b} of the master equation into a stochastic equation
for a pure state. In steady, state, the pure state will move
ergodically within some (perhaps infinite) ensemble of pure
states. This is how an unraveling defines an ensemble, with the
weighting of each member being the proportion of time the system
spends with that state.

\subsubsection{Survival Probability}

\label{secnotpur}

Third, in Ref.~\cite{WisVac98} we defined robustness in terms of
the fidelity or survival probability of the pure states rather
than their purity. That is, we consider how close the states
remain  to their original state under the master equation
evolution, rather than just how close they remain to a pure state.
This means that Hamiltonian evolution alone can affect the
robustness of states (whereas it does not affect their purity,
except in conjunction with the irreversible terms). It might be
thought that this is an undesirable feature. However, as will be
shown,  using the survival probability gives results that accord
with the usual concept of coherence in lasers. This contrasts with
the results that are obtained using purity, which we also consider
at the end of this paper (Sec.~V~C)

\subsubsection{Survival Time}
\label{secst}

The final aspect of our work that differs from  most previous
approaches \cite{Gea98,Gal95,ParScu98,DioKie00} is that we
quantify the robustness by the survival time. (This time was
previously called the fidelity time in Ref \cite{BarBurVac96}).
It is the time taken for the survival probability to fall below
some predefined threshold.  This is as opposed to considering the
rate of decay of the survival probability at the initial time.
That rate is actually identical to half the initial rate of decay
of the purity, and hence is independent of any Hamiltonian terms.
It is only by considering the robustness over some finite time
that the Hamiltonian terms will contribute.

\subsection{Unraveling the Master Equation}

In this section, we briefly reiterate the discussion in
Ref.~\cite{WisVac01a} on how the master equation is unraveled to
yield a pure state ensemble.  The most general form of the
Markovian master equation  is \cite{Lin76}
\beq
  \dot{\rho}= -i[H,\rho] + \sum_{k=1}^K {\cal D}[c_k]{\rho}
  \equiv {\cal L}\rho ,\label{genme}
\eeq
where for arbitrary operators $A$ and $B$,
\beq
  {\cal D}[A]B \equiv ABA\dg - \{A\dg A,B\}/2.
  \label{defcalD}
\eeq
We assume this to have a unique stationary state $\rho_{\rm ss}$.
It can be represented in terms of pure states as
\beq
  \rho_{\rm ss} = \sum_{n} \wp_{n} P_{n},
  \label{genpse}
\eeq
where the $P_{n}$ are projection operators and the $\wp_{n}$   are
positive weights summing to unity. The (possibly infinite) set of
ordered pairs,
\beq
  E = \{ ( P_{n},\wp_{n}) :n=1,2,\ldots \},
\eeq
we will call an ensemble $E$ of pure states. There are
continuously infinitely many ensembles $E$ that represent
$\rho_{\rm ss}$. Our aim is to find the `best' or `most natural'
representation for $\rho_{\rm ss}$.

Our first requirement is that the ensemble be physically
realizable. This is possible if the environment of the system is
monitored, leading to a stochastic quantum trajectory for the
system state. Assuming that the initial state of the system is
pure, the quantum trajectory for its projector will be described
by the stochastic master equation (SME)
\beq
  d{P}=  dt\left[ {\cal L} + {\cal U}(t)  \right]P.
  \label{SSE1}
\eeq
Here the superoperator ${\cal U}$, which we will call an
unraveling, does not affect the average evolution of the system,
but preserves the idempotency of $P$. In the long-time limit the
system will be in some pure state $P_{n}$, with some probability
$\wp_{n}$ such that \erf{genpse} is satisfied. Since the states
and weights will depend on the unraveling ${\cal U}$, we denote
the resultant stationary ensemble by
\beq
   E^{\cal U} = \{ (P_n^{\cal U}, \wp_n^{\cal  U}) :n=1,2,\ldots \}.
   \label{ens}
\eeq

For practical reasons explained in \cite{WisVac01a}, we restrict
our investigation of the (atom) laser to continuous Markovian
unravelings (CMUs). As was shown in \cite{WisVac01a}, under a
linearization of the dynamics these lead to Gaussian pure states
as the members of the ensembles $E^{\cal U}$. As mentioned above,
we will also consider ensembles, in particular Gaussian ensembles,
which are not constrained by the requirement of physical
realizability. This is in order to see the importance of this
requirement in constraining the most natural ensembles.

\subsection{Quantifying the Robustness}

\subsubsection{Survival Probability}

Imagine that the system has been evolving under a particular
unraveling ${\cal U}$ from an initial state at time $-\infty$ to
the stationary ensemble at the present time $0$. It will then be
in the state $P_n^{\cal U}$ with probability $\wp_n^{\cal U}$. If
we now cease to monitor the system then the state will no longer
remain pure, but rather will relax toward $\rho_{\rm ss}$ under
the evolution of Eq.~(\ref{genme}).

This relaxation to equilibrium will occur at different rates for
different states. For example, some unravelings will tend to
collapse the system at $t=0$ into a pure state that is very
fragile, in that the system will not remain in that state for very
long. In this case the ensemble would rapidly become a poor
representation of the observer's current knowledge about the
system. Hence we can say that such an ensemble is a `bad'   or
`unnatural' representation of   $\rho$. Conversely, an unraveling
that produces robust states would remain an accurate description
for a relatively long time. We expect such a `good' or `natural'
ensemble to give more intuition about the dynamics of the system.
The most robust ensemble we interpret as the `best' or `most
natural'   such ensemble.

In most of this paper we quantify the {\em robustness} of a
particular state $P_n^{\cal U}$ by its survival probability
$S_n^{\cal U}(t)$. This is the probability that the system would
be found (by a hypothetical projective measurement) to be still in
the state $P_n^{\cal U}$ at time $t$. It is given by
\cite{fidelity}
\beq
   S_n^{\cal U}(t) = {\rm Tr}[P_n^{\cal U}
   e^{{\cal L}t}P_n^{\cal U}].
   \label{overlap}
\eeq

Since we are considering an ensemble $E^{\cal U}$ we must define
the average survival probability
\beq
   S^{\cal U}(t) = \sum_{n} \wp_n^{\cal U} S_n^{\cal U}(t).
\eeq
In the limit $t\to\infty$ the ensemble-averaged survival
probability will tend towards the stationary value
\beq
   S^{\cal U}(\infty) = {\rm Tr}[ \rho_{\rm ss} ^2].
\eeq
This is independent of the unraveling ${\cal U}$ and is a measure
of the mixedness of $\rho_{\rm ss}$.

\subsubsection{Comparison with Purity}\label{compur}

As noted in Sec.~\ref{secnotpur} above, it is more common in
discussions of robustness to use purity rather than survival
probability. The purity of a state at time $t$ can be quantified
as
\beq
  p^{\cal U}_{n}(t) = {\rm Tr}[\left(
    e^{{\cal L}t}P_n^{\cal U}\right)^{2}].
   \label{defpur}
\eeq
The ensemble average of this quantity is also initially unity, and
approaches ${\rm Tr}[ \rho_{\rm ss} ^2]$ as $t\to \infty$.
Alternatively, the purity could be quantified as the maximum
overlap of any pure state $\tilde{P}_n(t)$ with the evolved
mixed state:
\beq
   \tilde{p}^{\cal U}_{n}(t)
   = {\rm max}_{\tilde{P}_n(t)}{\rm Tr}
   [\tilde{P}_n(t)\left(e^{{\cal L}t}P_n^{\cal U}\right)].
   \label{altpur}
\eeq
For Gaussian states (see Sec.~III) these quantities are simply
related by $\tilde{p}^{\cal U}_{n}(t)=2/[1+1/p^{\cal U}_{n}(t)]$.

The survival probability has a number of advantages over purity.
First, we motivated our robustness criterion from the desire for
$E^{\cal U}$ to {\em remain} a good description of the system once
the unraveling ceases. That is, we wish to be able to usefully
regard the members of the ensemble $E^{\cal U}$ as the states the
system is ``really'' in at steady state.  This is better
quantified by the survival probability because the purity
effectively takes into account only how close the state $e^{{\cal
L}t}P_n$ remains to some pure state $\tilde{P}_n(t)$ [introduced
in \erf{altpur}], not how close it remains to the original state
$P_n$. An ensemble constructed by considering the purity would
thus, in general, only remain a good description of the system by
including the deterministic (but not necessarily unitary)
evolution of its members from $P_n$ to $\tilde{P}_n(t)$ after the
unraveling ceases. This time evolution would negate the idea that
the ensemble of states $P_n$ is the best representation of the
system at steady state.

Another reason for preferring the survival probability comes from
imagining that the unraveling ${\cal U}$ continues after $t=0$. In
that case one can calculate a {\em conditional} survival
probability, being the overlap of the pure  conditional state with
the pure initial state. The ensemble average of this conditional
survival probability is simply the survival probability
$S_{n}^{\cal U}(t)$ defined in \erf{overlap} above. Thus the
concept of survival probability still applies even for the
conditional evolution. By contrast, the conditional purity of the
unraveled state would always be unity, and consequently has no
relation to the unconditional purity defined in \erf{defpur}. The
latter thus has no simple interpretation for the unraveled
evolution.

The final reason for preferring survival probability, already
noted in Sec.~\ref{secnotpur}, is that it yields results for the
atom laser that have a  clear and simple physical interpretation
in terms of the coherence of the laser output. We will show that
this is so in the Discussion section.

One limit in which quite different results are to be expected from
using purity rather than survival probability is that in which the
Hamiltonian part of the dynamics dominates. As will be shown, this
limit is highly relevant for the atom laser.

Formally, we split the Liouvillian superoperator ${\cal L}$ as
\beq
  {\cal L} = {\cal L}_{\rm irr} + \chi {\cal L}_{\rm rev},
\eeq
where $\chi$ is a large parameter and
\bqa
   {\cal L}_{\rm irr}\rho &=& \sum_{k=1}^{K}{\cal D}[c_{k}]\rho. \\
   {\cal L}_{\rm rev}\rho &=& -i[H,\rho].
\eqa
The reversibility of ${\cal L}_{\rm rev}$ implies that
\beq
    {\rm Tr}[A{\cal L}_{\rm rev}B] = -{\rm Tr}[B{\cal L}_{\rm rev}A]
\eeq
and so ${\rm Tr}[A{\cal L}_{\rm rev}A]=0$, for arbitrary operators
$A$ and $B$.

To first order in time, both the survival probability and the
purity depend only upon the irreversible term:
\bqa
      S(t) &=& 1 + t{\rm Tr}[P{\cal L}_{\rm irr}P] ,
                  \label{aasurshortt}\\
      p(t) &=& 1 + 2t{\rm Tr}[P{\cal L}_{\rm irr}P] .
\eqa
For longer times, both expressions will (in general) be dominated
in the large-$\chi$ regime by the reversible term, but in
different ways:
\bqa
  S(t) &\simeq& 1 + \chi^{2} (t^{2}/2)
             {\rm Tr}[P{\cal L}_{\rm rev}^{2}P],
      \label{aasurlongt}\\
  p(t) &\simeq& 1 + \chi t^{2} {\rm Tr}[P({\cal L}_{\rm irr}
            {\cal L}_{\rm rev} - {\cal L}_{\rm rev}{\cal L}_{\rm irr})P].
\eqa
The Hamiltonian term directly affects the survival probability,
but it affects the purity only in combination with the
irreversible term.

\subsubsection{Survival Time} \label{secsurvt}

The above analysis shows that the difference between purity and
survival probability only shows up at finite times. Thus the best
way to characterize robustness is to look not at the initial rate
of decay of the survival probability, but at the time it takes to
fall below some threshold value $\Lambda$ satisfying
\beq
   1>\Lambda > {\rm Tr}[\rho_{\rm ss}^{2}].
\eeq
The ensemble survival time  for a particular unraveling would then
be defined as
\beq
   \tau^{\cal U} = {\rm min} \{ t\,:\, S^{\cal U}(t)= \Lambda  \}.
   \label{deftau}
\eeq
Note that this time is the {\em first} time for which $S^{\cal
U}(t)= \Lambda$. The survival probability is not necessarily
monotonically decreasing and in some simple examples there will be
many solutions to the equation  $S^{\cal U}(t)= \Lambda$
\cite{WisBra00}.

A natural choice of $\Lambda$, suggested in Ref.~\cite{WisVac98},
is the maximum eigenvalue of $\rho_{\rm ss}$:
\bqa
     \Lambda &=& \lim_{n\to \infty}
            \left({\rm Tr}[\rho_{\rm ss}^{n}]\right)^{1/n}\\
            &=& {\rm max}\{\lambda_{j} \in I\!\!R \,:\, \rho_{\rm ss}Q_{j}
            = \lambda_{j} Q_{j}, Q_{j}= Q_{j}^{2}\}.
     \label{defLambda}
\eqa
This can be shown to satisfy $\Lambda > {\rm Tr}[\rho_{\rm
ss}^{2}]$ as follows. Let the eigenvalues of $\rho_{\rm ss}$ be
ordered such that $\Lambda=\lambda_{1} \ge \lambda_{2} \ge
\lambda_{3} \ldots$. Then
\bqa
  {\rm Tr}[\rho_{\rm ss}^{2}] &=& \Lambda^{2} + \sum_{j=2}
             \lambda_{j}^{2} \\
    &<& \Lambda^{2} + \sum_{j=2} \Lambda \lambda_{j} \\
    &=& \Lambda^{2} + \Lambda (1-\Lambda) = \Lambda.
\eqa
Here the strict inequality holds unless all eigenvalues of
$\rho_{\rm ss}$ are equal.

In the absence of any monitoring of the bath, the projector
$Q_{1}$ would be one's best guess for what pure state the system
is in at steady state. The chance of this guess being correct is
simply $\Lambda$, which is obviously independent of time $t$.
Using this $\Lambda$, the survival time $\tau^{\cal U}$ could thus
be interpreted as the time at which the initial state $P^{\cal
U}_{n}$ {\em  ceases (on average) to be any better than} $Q_{1}$
as an estimate of which pure state is occupied. In other words,
the ensemble $E^{\cal U}$ is obsolete at time $\tau^{\cal U}$.

In this paper we do not use this choice for $\Lambda$, for reasons
to be explained later. This brings a certain degree of
arbitrariness into the analysis. However, as we show, the most
important and interesting results we obtain are independent of the
choice of $\Lambda$.

Having chosen a particular value for $\Lambda$, the survival time
$\tau^{\cal U}$ quantifies the robustness of an unraveling ${\cal
U}$.  Let the set of all unravelings be denoted $J$. Then the
subset of {\em maximally robust} unravelings $J_{M}$ is
\beq
   J_{M} = \{ {\cal R} \in J : \tau^{\cal R} \geq \tau^{\cal U} \;
    \forall\, {\cal U} \in J \}.
\eeq
As noted above, in practice it may be necessary to restrict the
analysis to continuous Markovian unravelings $D$, and the
corresponding subset $D_{M}$. Even if $J_{M}$ has many elements
${\cal R}_{1}, {\cal R}_{2}, \cdots$, these different unravelings
may give the same ensemble $E^{\cal R}=E^{{\cal R}_{1}}=E^{{\cal
R}_{2}}=\cdots$. In this case $E^{\cal R}$ is the most natural
ensemble representation of the stationary solution of the given
master equation. When we consider ensembles that are not
constrained by the condition of physical realizability, we will
denote the most robust of these by $E^{\rm R}$. That is, we
reserve the calligraphic ${\cal R}$ to denote a robust unraveling.

\section{MRUs for The (Atom)    Laser}

\subsection{The Master Equation}

The master equation we use for the (atom) laser is the same as
that in the preceding paper \cite{WisVac01a}.  In    the
interaction picture, and measuring time in units of the output
decay rate, it  is
\bqa
   \dot\rho &=& \left( \mu  {\cal D}[a\dg] {\cal A}[a\dg]^{-1}
                  + {\cal D}[a] + N{\cal D}[a\dg a] \right)\rho \nn \\
           && -\, i C[(a\dg a)^{2},\rho].
   \label{lasme}
\eqa
The parameters $N$ and $C$ represent excess phase noise and
self-interaction energy respectively. This has the stationary
solution expressed in Eqs.~(\ref{mixnum}) and (\ref{mixcoh}), with
mean boson number $\mu$.

To make progress on this equation we linearized it around a mean
field by making the replacement
\beq
    a = \sqrt\mu +  (x + i y )/2,
\eeq
with $x$ and $y$ Hermitian. The linearized master equation has a
Gaussian solution with moments
\beq
    \mu_{mn} = \ip{(x^{m}y^{mn})_{\rm sym}}
\eeq
given by
\bqa
   \mu_{10}(t) &=& \mu_{10}(0) w ,\label{solns1}\\
   \mu_{01}(t) &=& \mu_{01}(0) - \chi \mu_{10}(0)(1-w),
                     \label{meanph}\\
   \mu_{20}(t) &=& \mu_{20}(0) w^{2} + 1 - w^{2}, \label{ampvar}\\
   \mu_{11}(t) &=& \mu_{11}(0) w - \chi\left\{ 1+ w[\mu_{20}(0) - 2]
             \right.\nl\left.  \phantom{\mu_{11}(0) w-\chi\left\{\right.}
             +\, w^{2}[1-\mu_{20}(0)] \right\} ,\\
   \mu_{02}(t) &=& \mu_{02}(0) + (2+\nu)t - 2\chi\mu_{11}(0)(1-w) \nl
                 + \, 2\chi^{2}\left\{ t + [\mu_{20}(0) - 2] (1-w)
                 \right.\nl\left.
                \phantom{2\chi^{2}\left\{\right.t}+\,  [1-\mu_{20}(0)]
                (1-w^{2})/2\right\},
                \label{solns5}
\eqa
where $w\equiv  e^{-t}$, $\chi=4\mu C$ and $\nu = 4\mu N$. The
long-time limit of this is a Wigner function
\beq
  W_{\rm ss}(x,y) \propto \exp(-x^{2}/2)
  \label{Wss}
\eeq
with amplitude quadrature ($x$) variance of unity and phase
quadrature ($y$) variance of infinity. This is what is expected as
the linearized version of the stationary state of \erf{mixnum}.

The conditions for the output of the laser to be coherent, in the
sense of having an atom flux much greater than the linewidth (as
conventionally defined) are simply stated in terms of the
dimensionless self-energy $\chi$ and excess phase diffusions $\nu$
\cite{WisVac01a}
\bqa
   \chi &\ll& \mu^{3/2}, \label{cohcon1} \\
   \nu &\ll& \mu^{2}.    \label{cohcon2}
\eqa

\subsection{The Unraveled Master Equation} \label{sec:pre}

Under a continuous Markovian unraveling the long-time solutions
for the linearized stochastic dynamics are still Gaussian
\cite{WisVac01a}. In fact, the evolution of the second order
moments $\mu_{20},\mu_{02},\mu_{11}$ is deterministic. This means
that for a given unraveling ${\cal U}$ the stationary ensemble
will consist of Gaussian pure states all having the same second
order moments. They are distinguished only by their first order
moments $\bar{x}=\mu_{10},\bar{y}=\mu_{01}$, which therefore take
the role of the index $n$ in \erf{ens}. The different ensembles
themselves are indexed by another pair of numbers,
$\mu_{11},\mu_{20}$, which play    the role of ${\cal U}$ in
\erf{ens}. We do not need $\mu_{02}$ because the purity of the
unraveled states implies that
\beq
   \mu_{20}\mu_{02} - \mu_{11}^{2} = 1.
   \label{pur}
\eeq
However, it should be noted that the mapping from ${\cal U}$ to
$\mu_{11},\mu_{20}$ is, in general, many-to-one.

The ensemble can thus be represented as
\beq
   E^{\cal U} = \{(P^{\cal U}_{\bar{x},\bar{y}},
          \wp^{\cal U}_{\bar{x},\bar{y}}) :{\bar{x},\bar{y}}\},
          \label{EcalU}
\eeq
where the second order moments of the pure state $P^{\cal
U}_{\bar{x},\bar{y}}$ are determined by the unraveling ${\cal U}$.

The weighting function is flat for $\bar{y}$ and for $\bar{x}$ is
given by \cite{WisVac01a}
\beq
   \wp^{\cal U}(\bar{x}) = [2\pi(1-\mu_{20})]^{-1/2}
       \exp\left\{ - \bar{x}^{2} / [2(1-\mu_{20})]\right\}.
       \label{Pbarx}
\eeq

It is convenient to use a new notation for the second order
moments,
\bqa
  \alpha &=& \frac{\mu_{02}}{\mu_{20}\mu_{02}-\mu_{11}^{2}} ,
    \label{alp}\\
  \beta &=& \frac{\mu_{11}}{\mu_{20}\mu_{02}-\mu_{11}^{2}} ,\\
  \gamma &=& \frac{\mu_{20}}{\mu_{20}\mu_{02}-\mu_{11}^{2}}
    \label{gam} .
\eqa
For pure states satisfying \erf{pur}, we have (as in the preceding
paper \cite{WisVac01a})
\beq
  \alpha = \mu_{02}\;;\;\;\beta = \mu_{11}\;;\;\; \gamma = \mu_{20},
\eeq
The different ensembles are now indexed by the pair
$\beta,\gamma$. Not all pairs   $\beta,\gamma$ correspond to
physically realizable ensembles. The method for determining which
do correspond to PR ensembles is described in the preceding paper
\cite{WisVac01a}, and the constraints that apply are simply
$\gamma > 0$ and
\beq
  (-2\chi\beta + 2+\nu)(-2\gamma+2) - (\beta+\chi\gamma)^{2} \geq 0.
  \label{PRC}
\eeq

\subsection{Survival Probability}

We are interested in the survival probability of the states
$P^{\cal U}_{\bar{x},\bar{y}}$. It is convenient to consider the
corresponding Wigner functions, $W^{\cal
U}_{\bar{x},\bar{y}}(x,y)$. Obviously the survival probability is
independent of $\bar{y}$ so  we will drop this subscript, and set
$\bar{y}=0$ for ease of calculation. For Gaussian states the
Wigner function is a bivariate Gaussian distribution with the
moments $\mu_{mn}$ defined above. The state with initial moments
$\mu_{mn}(0)$ will evolve into a state with moments $\mu_{mn}(t)$
given by (\ref{solns1}--\ref{solns5}). We will denote the Wigner
function for the former state    $W_{\bar{x}}(x,y,0)$ and that for
the latter $W_{\bar{x}}(x,y,t)$. The survival probability of the
state $P_{\bar x}^{\cal U}$ is given by \cite{Hil84} \singlecol
\bqa
  S_{\bar{x}}(t) &\equiv&
         {\rm Tr}[P_{\bar{x}}e^{{\cal L}t}P_{\bar{x}}]
      =  4\pi \int dx dy W_{\bar{x}}(x,y,0)W_{\bar{x}}(x,y,t) \\
     &=& 4\pi \int dx dy {\cal N}(0)
         \exp\left[\frac{\mu_{20}(0)\mu_{02}(0)}
         {\mu_{20}(0)\mu_{02}(0)-\mu_{11}(0)^{2}}
         \left(- \frac{(x-\bar{x})^{2}}{2\mu_{20}(0)} +
         \frac{\mu_{11}(0) (x-\bar{x})y}{\mu_{20}(0)\mu_{02}(0)} -
         \frac{y^{2}}{2\mu_{02}(0)}\right)\right] \\
     && \times \; {\cal N}(t) \exp\left[
         \frac{\mu_{20}(t)\mu_{02}(t)}{\mu_{20}(t)\mu_{02}(t)-\mu_{11}(t)^{2}}
         \left(- \frac{(x-\bar{x}w)^{2}}{2\mu_{20}(t)} + \frac{\mu_{11}(t)
         (x-\bar{x}w)(y+\chi\bar{x}(1-w))}{\mu_{20}(t)\mu_{02}(t)} -
         \frac{(y+\chi\bar{x}(1-w))^{2}}{2\mu_{02}(t)}\right)\right], \nn
\eqa
where
\beq
   {\cal N} = \left( 2\pi \sqrt{\mu_{20}\mu_{02}-\mu_{11}^{2}}
    \right)^{-1}.
\eeq

This survival probability should be averaged over all $\bar{x}$,
weighted by the distribution (\ref{Pbarx}) to get
\beq
   S^{\cal U}(t) = \int d\bar{x} S_{\bar{x}}(t)
                   \wp^{\cal U}(\bar{x}).
\eeq
Thus $S(t)$ is given by a triple Gaussian integral that evaluates
to the following:
\beq
   S^{\cal U}(t) = 2\sqrt{\frac{ (\alpha_{t}\gamma_{t} -
                   \beta_{t}^{2})/[1+ (1-\gamma_{0})R_{t} ]}
                   {(\alpha_0+\alpha_t)(\gamma_0+\gamma_t)-(\beta_0+\beta_t)^2}}
\eeq
where
\bqa
  R_t &=& \alpha_0+\alpha_tw^2+2\beta_t\chi wz+\gamma_t\chi^2 z^2
          -\frac{(\alpha_0+\alpha_tw+\beta_t\chi z)^2}{\alpha_0+\alpha_t}
          \nn \\
     & &\ \ \-\frac{\big[(\beta_0+\beta_t)(\alpha_0+\alpha_tw+\beta_t\chi z)
        -(\alpha_0+\alpha_t)(\beta_0+\beta_tw+\gamma_t\chi z)\big]^2}
        {(\alpha_0+\alpha_t)[(\alpha_0+\alpha_t)(\gamma_0+\gamma_t)-
        (\beta_0+\beta_t)^2]} ,
\eqa
\doublecol
where $z  \equiv 1-w $ and $\alpha,\beta,\gamma$ are as in
Eqs.~(\ref{alp})--(\ref{gam}), and $\mu_{mn}$ are as in
Eqs.~(\ref{solns1})--(\ref{solns5}). Note that at $t=0$ the state
is pure, so that
$\alpha_{0}=\mu_{02},\beta_{0}=\mu_{11},\gamma_{0}=\mu_{20}$ as
previously. The survival probability $S^{\cal U}(t)$ is thus a
function    of the initial state parameters $\gamma_{0}$ and
$\beta_{0}$, and the dynamical parameters $\nu$ and $\chi$.
%Because    $\mu_{20}$ cannot be greater than one (the stationary variance
%in $x$), we    also have the restriction $\gamma_{0} \leq 1$. Other
%restrictions on    $\gamma_{0}$ and $\beta_{0}$ come from the
%condition of physical realizability discussed in Sec.~\ref{sec:pre}.

\subsection{The Survival Time}

Following the general theory described in Sec.~\ref{secsurvt}, we
define the survival time $\tau^{\cal U}$ as the smallest (in this
case it will be the only) solution to the equation
\beq
  S^{\cal U}(\tau^{\cal U}) = \Lambda ,
\eeq
where $\Lambda$ is a constant satisfying
\beq
  1 > \Lambda > {\rm Tr}[\rho_{\rm ss}^{2}].
  \label{lammustsat}
\eeq

From the solution (\ref{rhoinf}) of    the nonlinear dynamics, the
lower bound on $\Lambda$ is, for $\mu \gg 1$,
\beq
   {\rm Tr}[\rho_{\rm ss}^{2}] = (4\pi\mu)^{-1/2}.
\eeq
In the same limit, the largest eigenvalue for $\rho_{\rm ss}$ is
\beq
  \lim_{n\to\infty}\sqrt[n]{{\rm Tr}[\rho_{\rm ss}^{n}]}
  = (2\pi\mu)^{-1/2}.
  \label{largeig}
\eeq

From these expressions it is evident that there  would be a
problem in choosing \erf{largeig} for $\Lambda$: it is very close
to the value for ${\rm Tr}[\rho_{\rm ss}^{2}] = (4\pi\mu)^{-1/2}$.
This means that the survival time would be equal to the time by
which the system has relaxed almost to the equilibrium mixed
state. In particular, its phase would necessarily be poorly
defined by this time, which means that the linearization of the
dynamics that we have been using would not be valid.

If instead we start with the solution (\ref{Wss}) of the
linearized dynamics, we have an even worse situation:
\beq
   {\rm Tr}[\rho_{\rm ss}^{2}]
   = \lim_{n\to\infty}\sqrt[n]{{\rm Tr}[\rho_{\rm ss}^{n}]}
   = 0.
\eeq
In this case the survival time would always be infinite, which is
not helpful.

Because of these problems, we have not chosen the largest
eigenvalue of $\rho_{\rm ss}$ for $\Lambda$. Instead we have
investigated the dependence of $\tau^{\cal R}$ on $\Lambda$ for
various values, namely $\Lambda = 0.5, 0.2, 0.1, 0.05$.  As
will be shown, the most robust ensemble, (that with the largest
survival time) is substantially independent of $\Lambda$. Unless
otherwise stated we choose $\Lambda$ to be the midpoint of the two
bounds in \erf{lammustsat}, namely
\beq
  \Lambda = 1/2.
\eeq

\subsection{Unconstrained Gaussian Ensembles}

Finding the most robust PR ensemble $E^{\cal R}$ consists of a
searching for the maximum $\tau$ in the region of $\beta$-$\gamma$
space allowed by the PR constraint. To determine how important
this constraint is in determining $E^{\cal R}$, we also search for
the maximum $\tau$ in all of $\beta$-$\gamma$ space (subject only
to $0< \gamma\leq 1$). The ensemble picked out by this search we
will call the most robust unconstrained ensemble and denote
$E^{\rm R}$. Although we call in unconstrained, it is in fact
constrained to be of the same form as the ensembles resulting from
a continuous Markovian unravelings. That is, it consists of
Gaussian states with identical second-order moments distinguished
only by their mean amplitude and phase.

\section{Results}

\subsection{Varying $\chi$ with $\nu=0$}
\label{svarchi} First we present the results for no excess phase noise
($\nu=0$) to see the effect of varying the self-energy parameter
$\chi$. Because our results are numerical, we present them mostly
in a graphical form.

\subsubsection{Evolution at $\chi = 0$ and $\chi=50$}
\label{chi0100}

Fig.~\ref{fig1} shows the evolution of various initially pure
Gaussian quantum states under the linearized evolution of
Eqs.~(\ref{solns1})--(\ref{solns5}). We represent these states
by the one-standard-deviation ellipses of the Wigner function. In
each case we choose the initial mean location of the state in
phase space to be $\bar{x}=\bar{y}=0$, and, for   the last two
cases, for $\bar{y}=0$, $\bar{x}=\pm \sqrt{3/2}$ as well.

The first case in Fig.~\ref{fig1}(a) is for $\nu=0,\chi=0$, and an
initial coherent state. The ellipses are plotted for $t=0,3,10$.
The middle time is the ensemble-averaged survival time for an
ensemble of coherent states; that is, the time at which the
ensemble-averaged survival probability $S(t)$ drops to $1/2$. For
the particular case of the coherent state there is no distinction
between the ensemble-averaged survival probability and the
survival probability of a single coherent state $S_{\bar x}(t)$.
That is because the $x$-variance $\gamma$ of a coherent state is
equal to unity, the ensemble-averaged  $x$-variance, so that
perforce $\bar{x}=0$. Note  that the only dynamics in evidence
here is phase diffusion, causing the $y$-variance of the state to
increase. For $\chi=\nu=0$, the coherent state ensemble is in fact
the most robust ensemble. This  can be verified analytically. It
is also physically realizable, as shown in the preceding paper
\cite{WisVac01a}.

The second case in Fig.~\ref{fig1}(b) is again for an initial
coherent state but with $\nu=0,\chi=50$, plotted for $t=0, 0.0678,
0.2$. Again the middle time is the survival time for the coherent
state. Note that it is almost  two orders of magnitude smaller
than the coherent state survival time for $\chi=0$. The effect of
the large $\chi$ is to rapidly shear the state. This is because
the ${a\dg}^{2}a^{2}$ nonlinearity amounts to an
intensity-dependent frequency shift. The coherent state ensemble
$E^{\ket{\alpha}}$, however, is not the most robust ensemble for
$\chi=50$.

The third case in Fig.~\ref{fig1}(c) is the most robust
unconstrained ensemble $E^{\rm R}$ for $\nu=0,\chi=50$, as
determined by the numerical method discussed in Sec.~III. Three
members, $\bar x=0, \pm \sqrt{3/2}$, of this ensemble are
displayed. Note that   the $t=0$ state is a highly
amplitude-squeezed state. In fact it is not {\em purely}
amplitude-squeezed; the $x$-$y$-covariance $\beta^{\rm
R}=\mu_{11}$ is equal to $0.225$. In general, the angle $\theta$
between the major axis of the ellipse and the $y$-axis is
\beq
  \theta = \frac{1}{2}\arctan\frac{2\beta}{\alpha-\gamma}
  = \frac{1}{2}\arctan\frac{2\beta\gamma}{1+\beta^{2}-\gamma^{2}}.
\label{theta}
\eeq
In the limit of small $\gamma$ and $\beta$ this becomes $\theta
\simeq \beta\gamma$. In this case, with $\gamma^{\rm R}=0.100$, we
have $\theta^{\rm R} = 1.2^{\circ}$. This angle of rotation is
almost too small to make out in the figure. It is nevertheless
interesting that this slight rotation persists for all $\chi > 0$,
and that it is actually in the opposite direction to the rotation
caused by the shearing. That is, as the most robust state evolves
it passes through a point where the squeezing is purely in the
amplitude. Because the $x$-variance $\gamma^{\rm R}$ of the states
in this ensemble $E^{\rm R}$ is  less than unity, the different
members of $E^{\rm R}$ have different  values of $\bar{x}$. The
three initial states we show, with $\bar{x}=0$ and $\bar{x}=\pm
\sqrt{3/2}$, are typical members of the ensemble. The states into
which these members of the most  robust ensemble evolve are
plotted for $t=0.100 =\tau^{\rm R}$ (the    survival time) and
$t=0.2$ [as in Fig.~\ref{fig1}(b)]. Note that the survival time is
significantly larger that that for    the coherent state ensemble
in Fig.~\ref{fig1}(b).

The final plot, Fig.~\ref{fig1}(d), shows typical members of the
most robust PR ensemble $E^{\cal R}$. That is, the most robust
ensemble that can be realized by unraveling the master equation.
It is very similar to the most robust unconstrained ensemble
$E^{\rm R}$, also being highly amplitude squeezed with
$\gamma^{\cal R} = 0.092$. The three times at which its evolution
is plotted are $t=0$, $t=0.098=\tau^{\cal R} $, and $t=0.2$ [as in
Figs.~\ref{fig1}(b) and \ref{fig1}(c)]. Note that the survival
time $\tau^{\cal R}$ is marginally smaller than that for the
unconstrained ensemble, $\tau^{\rm R}$. The principal difference
from Fig.~\ref{fig1}(c) is that the $x$-$y$-covariance has the
opposite sign, with $\beta^{\cal R} = -0.092$. This corresponds to
a rotation of $\theta^{\cal R}=-0.48^{\circ}$, a rotation that is
accentuated as the evolution progresses. Again, the initial
rotation is almost too small to see in the figure, but it is a
persistent feature for large $\chi$.

From Fig.~\ref{fig1} it is evident that the evolved states from
the initial state with $\bar{x}=0$ in the robust cases (c) at
$t=0.100$ and (d) at $t=0.098$ are much closer to the initial
state than   the evolved state in the coherent   case (b) is at
time $t=0.0678$. This is despite the fact that all of these times
are the respective survival times at which the survival
probability drops to $1/2$. However, the evolved states from the
initial states with $\bar{x}=\pm \sqrt{3/2}$ in cases (c) and (d)
have a lower overlap with their initial states than does the
evolved coherent state of case (b). This clearly illustrates that
the survival probability is necessarily a property of the whole
ensemble of states, not of a single member. Figure~\ref{fig1} also
shows that the survival probability decays for different reasons
in different cases. In case (a) it decays because the evolved
state becomes more mixed, due to phase diffusion. In case (b) it
decays primarily because the evolved state changes shape
(shearing) while remaining relatively pure. In cases    (c) and
(d) it decays substantially because the mean position   of the
evolved state moves away from that of the initial states in phase
space. In Fig.~\ref{fig2} we compare the ensemble-averaged
survival probability $S(t)$ for the four cases in Fig.~\ref{fig1}.
Note that the time scale for case (a) ($\chi=0$) differs from that
used for cases (b), (c) and (d) ($\chi=50$). For   short times the
survival probability for the coherent state ensemble
$E^{\ket{\alpha}}$ (b) is greater than the survival probability
for the most robust ensembles $E^{\rm R}$ (c) and $E^{\cal R}$
(d). Indeed, the gradient of the survival probability for the
coherent state ensemble at $t=0$ is much less than that of the
most robust ensembles. This underlines the importance of    the
survival time, rather than the initial rate of decay of survival
probability, to quantify robustness. At short times the survival
probability generally decays linearly, due to irreversible
processes, as discussed in Sec.~\ref{compur}. A coherent state
minimizes this form of decoherence, resulting in an almost
quadratic behaviour of $S^{\ket{\alpha}}(t)$ for $t \alt \chi^{-1}
= 0.02$. This can be understood from the asymptotic analytical
expression in \erf{aasurlongt} for the survival probability for a
master equation with a large reversible term. This expression only
applies for the survival probability of a single state, but is
applicable to a coherent state ensemble because all members are
effectively identical. It need not, and indeed does not, apply to
the more robust ensembles. In comparison with the coherent state
ensemble, the most robust ensembles are affected more   by
irreversible evolution at short times but less by the interplay of
reversible and irreversible terms at longer times.

\subsubsection{Most Robust Unconstrained Ensemble
 for varying $\chi$.}\label{mruchi}

Having looked in detail at $\chi=0$ and $\chi=50$ we now present
an overview for $\chi$ ranging from $1$ to $10,000$. In this
section we concentrate upon the most robust unconstrained
ensemble. In Fig.~\ref{fig3} we plot the second-order moments
$\alpha^{\rm R},\beta^{\rm R},\gamma^{\rm R}$ defining the most
robust unconstrained ensemble $E^{\rm R}$,    as a function of
$\chi$. We also plot the survival time $\tau^{\rm R}$ for this
ensemble, and, for comparison, the survival time
$\tau^{\ket{\alpha}}$ for an ensemble consisting of coherent
states.

For values of $\chi$ less than about $7.7$, the members of the
most robust unconstrained ensemble are close to coherent states,
with $\alpha^{\rm R} \approx \gamma^{\rm R} =1$ and $\beta^{\rm R}
\alt 1$. As noted above, the states are sheared in the {\em
opposite} direction to the shearing produced by $\chi$. At $\chi
\approx 7.7$ there is a discontinuity in all state parameters.
Below this value the maximum survival time $\tau$ lies on the
boundary $\gamma=1$. Above this value, what was previously a local
maximum at some point $\gamma < 1$ becomes a global maximum, hence
the jump in the parameters. This is shown by the contour plots of
$\tau$ versus $\gamma$ and $\beta$ in Fig.~\ref{fig4}.

As $\chi$ becomes large, all of the curves plotted in
Fig.~\ref{fig3} tend    to straight lines on the log-log plot. It
is thus an easy matter to read off the following power laws from
the gradients of these lines:
\bqa
   \alpha^{\rm R}  &\sim& \chi^{2/3} ,
                        \label{scal1} \\
   \beta^{\rm R} &\sim& \chi^{-1/3} ,  \\
   \tau^{\rm R} \simeq \gamma^{\rm R} &\sim&   \chi^{-2/3} .
                         \label{scal4}
\eqa
These results clearly show that as $\chi$ increases, the most
robust states become increasingly amplitude-squeezed. From
\erf{theta} the scaling law for the rotation angle of the squeezed
state is
\beq
   \theta^{\rm R}  \sim \chi^{-1}.
\eeq

These scalings with $\chi$ can be understood by considering the
causes of the decay in the survival probability from the equations
(\ref{solns1})--(\ref{solns5}). A typical highly
amplitude-squeezed state member of the most robust ensemble has a
mean amplitude-quadrature fluctuation $\bar{x}$ of order unity.
From \erf{meanph}, the mean $y$-quadrature will therefore change
in a time $t  \ll 1$ by an amount of order $\chi t$.  This will
result in the significant decay of the survival probability if the
change $\chi t$ is of order the standard deviation $\alpha^{1/2}$
of the $y$-quadrature for that squeezed state; in other words, if
$t=\tau$ where
\beq
   \tau \sim \alpha^{1/2}\chi^{-1}.
   \label{anscal1}
\eeq
This reduction in overlap due to the motion of the mean phase of
the states is clearly illustrated in Fig.~\ref{fig1}(c) for the
initial states with $\bar{x}=\pm \sqrt{3/2}$. The survival
probability will also be affected by an increase in the phase
quadrature variance $\mu_{02}$. From \erf{solns5}, the dominant
terms for short times are $\mu_{02}(t)-\alpha = -2 \chi \beta t
+\chi^2 \gamma t^2$. Evidently a positive value of the initial
$x$-$y$ covariance $\beta$ can, at some time $t$, cancel the
increase in the phase variance caused by the nonzero initial
amplitude variance $\gamma$. This effect will maximize the
survival probability if the cancellation occurs at a time of order
the survival time $\tau$. This gives the second condition
\beq
   \tau \sim \gamma^{-1} \beta \chi^{-1} .
   \label{anscal2}
\eeq
This effect is most easily seen for the $\bar{x}=0$ initial state
in Fig.~\ref{fig1}(c), where the phase variance at the survival
time is little changed from its initial value whereas the phase
variance a short time later is significantly changed. Lastly, we
consider the effect of motion and diffusion in the $x$ direction.
From \erf{ampvar}, the amplitude-quadrature variance increases at
a rate of order unity. It  will cause a drop in the survival
probability once the increase is comparable to the initial
amplitude variance $\gamma$, which is at $\tau \sim \gamma$. From
\erf{solns1} the mean amplitude $\bar{x}$ decays to 0 at rate
unity, but this will only cause a significant drop in $S(\tau)$
when the decrease in amplitude is of the order of the amplitude
standard deviation, that is for $\tau \sim \gamma^{-1/2}$, which
is much longer. Thus the third condition is just
\beq
   \tau \sim \gamma \sim \alpha^{-1}.
   \label{anscal3}
\eeq
Once again, the $\bar{x}=0$ initial state in Fig.~\ref{fig1}(c)
shows that there is indeed a significant increase in the amplitude
variance at $t$ equal to the survival time.

The maximum survival time will clearly be when the survival times
from the effects above which cause decay of the survival
probability are comparable. The unique solutions to the three
analytical scaling relations (\ref{anscal1})--(\ref{anscal3}) are
the scaling laws found numerically and given in equations
(\ref{scal1})--(\ref{scal4}) above.

Not only does  $\tau^{\rm R}$ scale in  the same way as
$\gamma^{\rm R}$, it actually asymptotes to $\gamma^{\rm R}$ for
large $\chi$. This is a consequence of our choice $\Lambda=1/2$,
as will be shown later. In any case, the ensemble-averaged
survival time clearly decreases with $\chi$, so that the
nonlinearity causes a loss of robustness in the system even under
a maximally robust unraveling. However, this loss of robustness is
much worse for other ensembles. For example, the coherent state
ensemble $E^{\ket{\alpha}}$ has a survival time that varies as
\beq
  \tau^{\ket{\alpha}} \sim \chi^{-1},
  \label{taualph1}
\eeq
as shown by the dash-dot-dot curve in Fig.~\ref{fig3}. Thus for
large $\chi$ the description of the laser steady state in terms of
the highly amplitude-squeezed states of the most robust ensemble
is much more useful than the conventional coherent state
description.

The scaling in \erf{taualph1} can be easily derived from
\erf{solns5}. Even more simply, it can in fact be derived from the
asymptotic analytical formula in \erf{aasurlongt} for the survival
probability for a master equation with a large reversible term.
With $P$ a coherent state with $\bar{x}=0$ and ${\cal L}_{\rm
rev}\rho=-i[(\chi/4)x^{2},\rho]$ we find for the solution
$S(\tau)=1/2$,
\beq
  \tau = \sqrt{8}\,\chi^{-1}.
\eeq
Even the  coefficient here is a reasonable approximation, as
Fig.~\ref{fig3} shows.

\subsubsection{Most Robust Physical Realizable Ensemble for varying
$\chi$}

Having examined the most robust unconstrained ensemble, we now
determine the effect of the physical realizability constraint as
$\chi$ varies from $1$ to $10,000$. This is shown in
Fig.~\ref{fig5}. It can be seen from this plot that the ensemble
parameters differ from those in Fig.~\ref{fig3} for all $\chi$.
That is, the PR constraint is active for all $\chi$. There is no
discontinuity in the parameters, because \erf{PRC} keeps the state
away from the maximum of $\tau$ in $\beta$--$\gamma$ space. This
is illustrated clearly in Fig.~\ref{fig4}, where the shaded
regions represent the PR states. It is also clear from
Fig.~\ref{fig4} that, for large $\chi$, $\beta$ is effectively
constrained to be negative, which is why  we plot $\pm\beta$
rather than just $\beta$ in Fig.~\ref{fig5}. That is, the shearing
is in the direction induced by the nonlinearity, rather than in
the opposing direction as adopted by an unconstrained ensemble.
The PR ensemble is, not surprisingly, more physically reasonable.

Despite these differences, the  scaling laws for $\alpha^{\cal
R}$, $|\beta^{\cal R}|$, $\gamma^{\cal R}$ and $\tau^{\cal R}$ are
the same for the most robust PR ensemble $E^{\cal R}$ as for the
most robust unconstrained ensemble, that is
\bqa
   \alpha^{\cal R} &\sim& \chi^{2/3} \\
   -\beta^{\cal R} &\sim& \chi^{-1/3} \\
   \tau^{\cal R} \sim \gamma^{\cal R} &\sim& \chi^{-2/3}.
\eqa
The scalings for $\alpha^{\cal R}$, $\gamma^{\cal R}$ and
$\tau^{\cal R}$ can be derived using the same reasoning as in the
preceding case. The scaling for $\beta^{\cal R}$ arises as
follows. For robustness the system would like to have $\beta$
positive, as argued above. The constraint forces it to be
negative, which is why $E^{\cal R}$ is always constrained, and is
situated on the boundary of the PR region in $\beta$--$\gamma$
space. For $\chi$ large and $\gamma$ small, the boundary of the PR
region can be found from \erf{PRC} to be
\beq
  -\beta = \chi\gamma^{2}/4 ,\label{quadcon}
\eeq
which here scales as $\chi^{-1/3}$.

\subsection{Varying $\nu$}
\label{svarnu}

We turn now to the effect of excess phase noise $\nu$. Figure
\ref{fig6} is an overview of the most robust PR ensemble for
$\chi=0$ and for $\nu$ ranging from $1$ to $10,000$. The behaviour
is very simple.  For $\nu \alt 2.3$ the most robust states are
coherent states. As $\nu$ increases they become increasingly
squeezed states. For all values of $\nu$ we have $\beta=0$ (which
is therefore not plotted), indicating that the most robust states
are purely amplitude-squeezed. The scaling laws derived from this
plot are
\bqa
  \alpha^{\cal R} &\sim& \nu^{1/2} , \\
  \gamma^{\cal R} &\sim& \nu^{-1/2},  \\
  \tau^{\cal R} \simeq \gamma^{\cal R} &\sim& \nu^{-1/2} .
\eqa
This ensemble is not constrained by the PR constraint (\ref{PRC}).
These scaling can again be deduced by arguments similar to those
in Sec.~\ref{mruchi}.  Unlike the nonlinear $\chi$ term, phase
diffusion does not cause motion of the mean position of a typical
squeezed state. Rather, from \erf{solns5}, it simply causes the
phase-quadrature variance to increase linearly as $\nu \tau$. The
survival probability will drop significantly in this time if $\nu
\tau$ is comparable to the original phase variance, $\alpha$. From
the increase in the amplitude variance we get $\tau \sim \gamma
\sim \alpha^{-1}$ as in Sec.~\ref{mruchi}. The maximum survival
time occurs when these two times are comparable, giving
$\tau^{\cal R} \sim \nu^{-1/2}$ and $\alpha^{\cal R} \sim
\nu^{1/2}$, as found numerically.

The survival time decreases with increasing $\nu$, and, once
again, it asymptotes to $\gamma^{\cal R}$ for large $\nu$. For
comparison we also plot the survival time $\tau^{\ket{\alpha}}$
for a coherent state ensemble. This scales as
\beq
  \tau^{\ket{\alpha}} \sim \nu^{-1},
\eeq
so that for large $\nu$ the most robust ensemble is much more
robust than the coherent state ensemble. This scaling can be
derived from the short time asymptotic analytic expression in
\erf{aasurshortt}. Since the excess phase diffusion dominates the
evolution for $\nu$ large we have approximately
\beq
  S(\tau) \simeq 1 + \nu t{\rm Tr}\{ P{\cal D}[x/2]P\}.
\eeq
Again, this expression only applies for a single state or an
ensemble such as the coherent state ensemble where all members are
effectively identical.  In the latter case it evaluates simply to
$1-\nu t/4$.

\subsection{Varying $\Lambda$}

The final parameter we wish to consider varying is $\Lambda$,
which defines the survival time $\tau$ by the equation
$S(\tau)=\Lambda$. All of the results presented so far were for
$\Lambda=0.5$. In Fig.~\ref{fig8} we show the parameters
$\alpha^{\cal R}$ and $\tau^{\cal R}$ for the most robust ensemble
as a function of $\chi$ for $\nu=0$ and for four values of
$\Lambda$.  For large $\chi$ the slope of the curves are
independent of $\Lambda$. Thus the scaling laws established in
Sec.~\ref{svarchi} are independent of $\Lambda$. As $\Lambda$
decreases, the survival time $\tau^{\cal R}$ increases, because it
takes longer for the survival probability to decay to that level.

Decreasing $\Lambda$ also causes the phase variance $\alpha^{\cal
R}$ to increase, indicating that the most robust states are more
highly squeezed. This is not unexpected, since the difference
between the coherent state ensemble and the most robust ensemble
is expected to be greater at longer times by the argument in
Sec.~\ref{chi0100}. However, the relative increase in
$\alpha^{\cal R}$ is far less than the relative increase in
$\tau^{\cal R}$. In other words, the most robust ensemble is only
weakly dependent on $\Lambda$. Interestingly, because
$\gamma^{\cal R} \sim 1/\alpha^{\cal R}$, $\gamma^{\cal R}$
decreases as $\Lambda$ decreases, while $\tau^{\cal R}$ increases.
Thus the asymptotic result $\gamma^{\cal R} \simeq \tau^{\cal R}$
can only be true at one value of $\Lambda$, namely $\Lambda=1/2$.

Figure \ref{fig9} presents the same information as Fig.~\ref{fig8}
but for $\chi=0$ and varying $\nu$ and $\Lambda$. Once again the
scaling laws established in Sec.~\ref{svarnu} are found to be
independent of $\Lambda$, and in this case the different values
for $\alpha^{\cal R}$ appear to asymptote. In this case, the value
for $\nu$ above which the coherent state ensemble ceases to be the
most robust ensemble increases for decreasing $\Lambda$. Above
these values of $\nu$ the amplitude-squeezing in the most robust
ensemble is always decreased as $\Lambda$ is decreased. However,
the difference is small (and may vanish as $\nu \to \infty$), so
that the equation $\tau^{\cal R} \simeq \gamma^{\cal R}$ is again
valid only for $\Lambda = 1/2$. The sum of these results justifies
our use of the single value $\Lambda=1/2$ for most of this work.

\section{Discussion}

\subsection{Summary}

The atom laser is an open quantum system with rich dynamics. In
this paper we have explored a new way of characterizing those
dynamics: finding the maximally robust unraveling \cite{WisVac98}.
This yields the most robust {\em physically realizable} ensemble
$E^{\cal R}$ of pure states $P^{\cal R}$ that survive the best. By
``surviving'', we mean remaining unaffected by the system
dynamics. This ensemble is, we have argued, the most natural
representation of the stationary state matrix $\rho_{\rm ss}$ of
the laser; if one wished to regard the laser as being ``really''
in a pure state, then the most natural states to choose are the
members of this ensemble. Although it is a time-independent
ensemble, it is drastically affected by alterations in the
dynamics of the atom laser that do not change the stationary state
matrix.

We considered a simple model for the atom laser in which
$\rho_{\rm ss}$ is a Poissonian mixture of number states of mean
$\mu$. Working in the linearized regime, we identified two
relevant dynamical parameters that may be varied without altering
this stationary state. The first is $\chi$, which is proportional
to the strength of self-interaction of the atoms in the laser. The
second is $\nu$, which is proportional to the excess phase
diffusion of the laser above the standard quantum limit.

For $\chi=0$ and $\nu$ small, the most robust ensemble was found
to consist of coherent states, with mean boson number $\mu$ but
with all possible phases. This is the most common representation
of the state of an optical laser, and so it is not surprising. In
terms of the parameters we used in the paper, the ensemble
consists of Gaussian pure states with phase quadrature variance
$\alpha=1$, amplitude-quadrature variance $\gamma=1$, and
amplitude--phase covariance $\beta = 0$.

As the self-energy $\chi$ is increased the most robust states
cease to be coherent states. In fact, for any nonzero value of
$\chi$, not only are the coherent states not the most robust
state; in addition they are not even physically realizable
\cite{WisVac01a}. For large values of $\chi$ the most robust
states $P^{\cal R}$ are very highly amplitude-squeezed states with
amplitude-quadrature variance $\gamma^{\cal R}$ scaling as
$\chi^{-2/3}$ and phase quadrature variance $\alpha^{\cal R}$
scaling as $\chi^{2/3}$. The same effect occurs for large values
of $\nu$, with scalings of $\nu^{-1/2}$ and $\nu^{1/2}$
respectively.

It is not known what value of $\nu$ would be appropriate to model
a realistic atom laser. However, it was argued in
Ref.~\cite{WisVac01a} that a typical value for $\chi$ might be
$1000$. This implies that the most natural description of an atom
laser would be in terms of highly amplitude squeezed stated, with
the standard deviation in the amplitude quadrature  being of order
$0.1$. Excess phase noise would only increase the amount of
squeezing in the states in the most robust ensemble.

As noted above, our analysis was based on a linearized
approximation for the laser dynamics. This is only valid if the
states under consideration have a well-defined coherent amplitude.
As $\chi$ or $\nu$ are increased indefinitely and the most robust
states become more amplitude-squeezed, this approximation will
clearly break down. Specifically, it will break down when the
phase variance predicted by the linearized analysis is of order
unity; that is, when the the phase quadrature variance
$\alpha^{\cal R}$ is of order the mean boson number $\mu$. From
the above scalings, for the linearization to remain valid we
require
\bqa
  \chi &\ll& \mu^{3/2},\label{wdcacon1} \\
   \nu &\ll& \mu^{2} \label{wdcacon2}.
\eqa

Although we cannot say with confidence what the most robust states
are when the linearization breaks down, we do know that they must
be states without a well-defined coherent amplitude (because that
is why the linearization breaks down). Therefore the conditions in
Eqs.~(\ref{wdcacon1}) and (\ref{wdcacon2}) also represent the
conditions for the most robust states to be states with
well-defined coherent amplitudes. In other words, if and only if
these conditions are satisfied, the most natural description of
the atom laser is in terms of states with a mean field.

\subsection{Interpretation}

We can now finally state the most important result of this paper.
The conditions (\ref{wdcacon1}) and (\ref{wdcacon2}) are identical
to the previously stated conditions (\ref{cohcon1}) and
(\ref{cohcon2}) for the output of the device to be coherent. Here
we mean coherent in the sense that the output is quantum
degenerate, with many bosons being emitted per coherence time.
Without this condition the device could not be considered a laser
at all, as its output would consist of independent atoms rather
than a matter wave.

The significance of this result is that {\em there is a perfect
correspondence between the `best' pure states for describing the
laser, and the coherence of its output.} If the most robust states
have a well-defined coherent amplitude, like coherent states, then
the output is coherent. If the most robust states do not have a
well-defined coherent amplitude, like number states, then the
output is not coherent. This profound result establishes
 the usefulness of maximally robust unravelings as an
investigational tool for open quantum systems.

It must be emphasized that the link between the presence or
absence of a mean field inside the laser, and the presence or
absence of quantum coherence in the laser output, is not due to
any simple relationship of definitions. Finding the maximally
robust ensemble is, as the diligent reader will appreciate, a very
involved process completely different from calculating the
first-order coherence function. In particular, the average
survival time for the members of the most robust ensemble has in
general no relationship with the coherence time.

\subsection{Comparison with Purity}

It is worth pointing out that the relationship we have established
between robust mean-field states and quantum degeneracy would not
have been found had we used purity rather than survival
probability as the basis of our definition for the most robust
ensemble. Although there are no great differences between the two
definitions as one varies $\nu$, there is a great difference as
one varies $\chi$. This is to be expected from the analysis in
Sec.~\ref{compur}, as $\chi$ scales the self-energy Hamiltonian,
whereas $\nu$ represents irreversible phase diffusion.

To prove this point we have calculated the ensemble that maximizes
the time it takes for the average purity of the member states [as
defined in \erf{defpur}] to drop to $1/2$ under the master
equation evolution. We plot  the parameters for this ensemble as a
function of $\chi$ in Fig.~\ref{fig9}. For comparison we also plot
the phase quadrature variance $\alpha^{\cal R}$ and the survival
time $\tau^{\cal R}$ of the most robust ensemble as previously
defined, in terms of survival probability. The ensemble parameters
when we use purity obey scaling laws for large $\chi$, but they
are different from those scaling laws obtained when using the
survival probability (Sec.~V~A~2):
\bqa
  \alpha^{\cal R'} &\sim& \chi^{1/2} , \label{scalp1} \\
  \beta^{\cal R'} &\approx& -1/4 ,\\
  \tau^{\cal R'} \;\sim\; \gamma^{\cal R'} &\sim& \chi^{-1/2}.
             \label{scalp4}
\eqa
As expected from Sec.~\ref{compur}, the purity half-life is much
longer than the survival time for large $\chi$. Here we use ${\cal
R}'$ rather than ${\cal R}$ to emphasize that we are using a
different measure of robustness.

The scalings in Eqs.~(\ref{scalp1})--(\ref{scalp4}) can be derived
analytically. For Gaussian states with moments $\mu_{mn}(t)$, the
purity at time $t$ is given by
\beq
  \tr{\rho^{2}(t)}
  = p(t)
  = [\mu_{20}(t)\mu_{02}(t)-\mu_{11}^{2}(t)]^{-1/2}.
\eeq
For $\nu=0$, $\gamma\ll 1$, $\beta\sim 1$, $\chi\gg 1$ and $t \ll
1$, as appropriate here, the solutions
(\ref{ampvar})--(\ref{solns5}), together with the condition
$p(t)=1/2$, yield the following equation for $\tau$
\beq
  3 \approx 2(1+\beta^{2})\tau/\gamma -2\chi\beta \tau^{2} + 2\chi^{2}\gamma
      \tau^{3}/3 + \chi^{2}\tau^{4}/3.
\eeq
It is clear from the term $O(\tau^{4})$ that $\tau$ will scale as
$\chi^{-1/2}$. To maximize $\tau$, the terms $O(\tau)$ and
$O(\tau^{3})$ imply that $\gamma$ should scale as $\chi^{-1/2}$ in
accord with \erf{scalp4} . The terms $O(\tau)$ and $O(\tau^{2})$
then imply that $\beta$ should be positive, and of order unity.
Indeed, for the unconstrained Gaussian ensemble $E^{R'}$ we find
$\beta \approx 1.8$. With the constraint of \erf{quadcon}, we get
$\beta$ negative and of order unity, as stated above.

The condition for the best purity-preserving states to have a
well-defined coherent amplitude is $\alpha^{\cal R'} \ll \mu$,
which from \erf{scalp1} gives
\beq
   \chi \ll \mu^2.
\eeq
This implies that there is a range of interaction strengths $
\mu^{3/2} \alt \chi \ll \mu^2$ for which the purity analysis
delivers a description of the laser in terms of states with a mean
field even though the laser output is no longer coherent in the
sense defined above. This regime can be interpreted in terms of a
the non-standard concept of {\em conditional coherence}, explored
in detail in the preceding paper \cite{WisVac01a}. The basic idea
is well illustrated by Fig.~\ref{fig1}. If one knows the mean
amplitude of the state with an uncertainty much less than unity,
as in Fig.~\ref{fig1}(d), then the direction that it will move in
phase space can be predicted with accuracy. This motion (which
amounts to different frequencies) can then be taken into account
in experiments  the output. Thus the spread in frequencies due to
spread in amplitude can be compensated for (up to a point).

\subsection{Comparison with Quantum State Diffusion}

A particular PR ensemble of interest is that generated by the
unraveling known as quantum state diffusion (QSD)
\cite{GisPer92,Dio88}. This is merely a particularly simple and
natural type of continuous Markovian unraveling. It has been
suggested \cite{DioKie00} that the corresponding ensemble is a
good candidate for the most robust ensemble. We investigated this
ensemble in the preceding paper \cite{WisVac01a} and found
analytically that its parameters $\beta$ and $\gamma$ have exactly
the same scaling as the PR ensemble $E^{\cal R'}$ based on
maximizing the robustness as measured by purity. That is, with
$\chi$,
\bqa
  \alpha^{\rm QSD} &\simeq&  \sqrt{2} \chi^{1/2}, \\
  \beta^{\rm QSD} &\simeq&  -1,\\
    \gamma^{\rm QSD} &\simeq& \sqrt{2} \chi^{-1/2}.
\eqa
and with $\nu$,
\bqa
  \alpha^{\rm QSD} &\simeq&  \frac{1}{\sqrt{2}} \nu^{1/2}, \\
  \beta^{\rm QSD} &=&  0 ,\\
    \gamma^{\rm QSD} &\simeq& \sqrt{2} \nu^{-1/2}.
\eqa
Consequently, the QSD ensemble $E^{\rm QSD}$ scales with $\chi$
quite differently from the maximally robust ensemble $E^{\cal R}$
according to our definition based on maximizing the survival time.
Thus unlike $E^{\cal R}$, but like $E^{\cal R'}$, the coherence of
its members does not have a direct correspondence with the laser
output coherence (in the conventional, unconditional, sense).

The correspondence (at least in scaling laws) between $E^{\rm
QSD}$ and $E^{\cal R'}$ is actually in contrast to the result
found by Di\'osi and Kiefer (for a different system)
\cite{DioKie00}. They found that PR states minimizing the loss of
purity were different from states produced by QSD. However, as
noted earlier, they considered only the initial rate of loss of
purity, which is insensitive to Hamiltonian terms. If they had
considered maximizing the half-life of the purity, as we have,
they may have obtained a different result.

\subsection{Future Work}

There are at least three future directions for this work. First,
the insights into the atom laser that the maximally robust
unravelings analysis offers suggests that this technique could be
applied fruitfully to other open quantum systems. It has already
been applied to fluorescent atoms \cite{WisBra00}, and could also
be applied to other quantum optical systems \cite{Car93b}, and
other models for Bose-Einstein condensates in equilibrium with a
reservoir \cite{Ang97}. These are all systems with nontrivial
dynamics, which could be more fully appreciated by determining the
maximally robust unraveling.

Second, the difference between the analyses based on survival
probability and purity deserves further investigation. As we
showed, the purity analysis gives a description of the laser mode
in terms of states with a well-defined coherent amplitude for high
values of $\chi$ where the survival analysis does not, and where
the output is not coherent in the conventional sense. Nevertheless
the results do make sense in terms of {\em conditional coherence}
\cite{WisVac01a}. Perhaps it is because purity is unaffected by
the motion of the mean position of the states in phase space that
it reflects conditional coherence, which relies on knowledge of
that motion to define the output mode. By contrast, the survival
probability {\em is} affected by the motion of the states, and
hence reflects conventional coherence that averages over the
different frequencies of rotation.

Finally, there are other approaches to quantifying the robustness
of unravelings apart from the survival probability and the purity.
For example, one could measure how quickly the unraveling purifies
the state, or how sensitive the purity is to imperfections in the
unravelings. Related ideas have recently been explored
\cite{DalDziZur01,Bra00}. These ideas could be best investigated
in systems somewhat simpler than the atom laser we have considered
here. This would give an indication for the robustness of the idea
of robustness; that is, how sensitive the maximally robust
unraveling is to the definition of robustness used, and which
definitions agree.

To conclude, the clear and simple interpretation for the results
we have obtained here for the atom laser vindicates our conviction
\cite{WisVac98} that maximally robust unravelings will have an
increasing role as a tool for understanding the dynamics of open
quantum systems.

\acknowledgments

J.A.V. thanks Profs. S.M. Barnett and K. Burnett for initial
discussions. H.M.W. is supported by the Australian Research
Council.

\begin{figure}
\psfig{figure=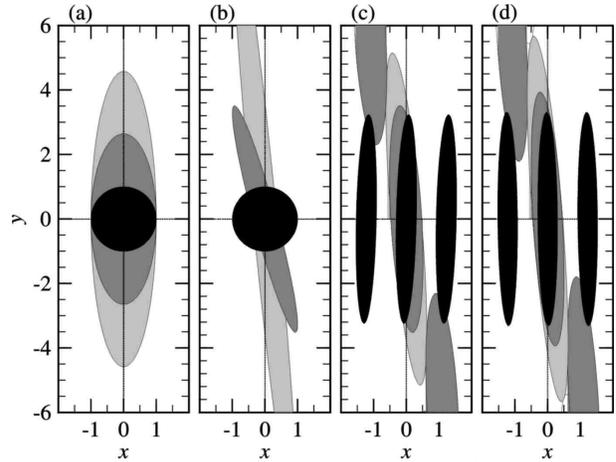,width=80mm,angle=-90,bbllx=7mm,bblly=7mm,bburx=200mm,bbury=258mm,clip=}
\caption{\narrowtext The evolution of  (initially pure) Gaussian
quantum states under the linearized laser master equation  for
four different cases. The states are represented by the one
standard-deviation ellipse of the Wigner function. In the all
cases we choose the initial mean location of the state in phase
space to be $\bar{x}=\bar{y}=0$, and for the last two we
additionally have $\bar{x}=\pm\sqrt{3/2}$. For all four cases the
excess phase diffusion is $\nu=0$. For case (a) we have $\chi=0$
and an initially coherent state (which forms the most robust
ensemble in this case). For case (b) we have $\chi=50$ and again
an initially coherent state. For case (c) we have
 $\chi=50$ but the initial states are members of
the most robust unconstrained ensemble $E^{R}$ for this $\chi$.
For case (d) we have  $\chi=50$ but the initial states are members
of the most robust PR-constrained ensemble $E^{\cal R}$ for this
$\chi$. In all cases the black ellipses are for $t=0$, the dark
grey ellipses for $t=\tau$ (the appropriate ensemble-averaged
survival time), and the light grey ellipses for a still later
time. Details of these times are given in the main text.
    \protect\label{fig1}}
\end{figure}

\begin{figure}
\psfig{figure=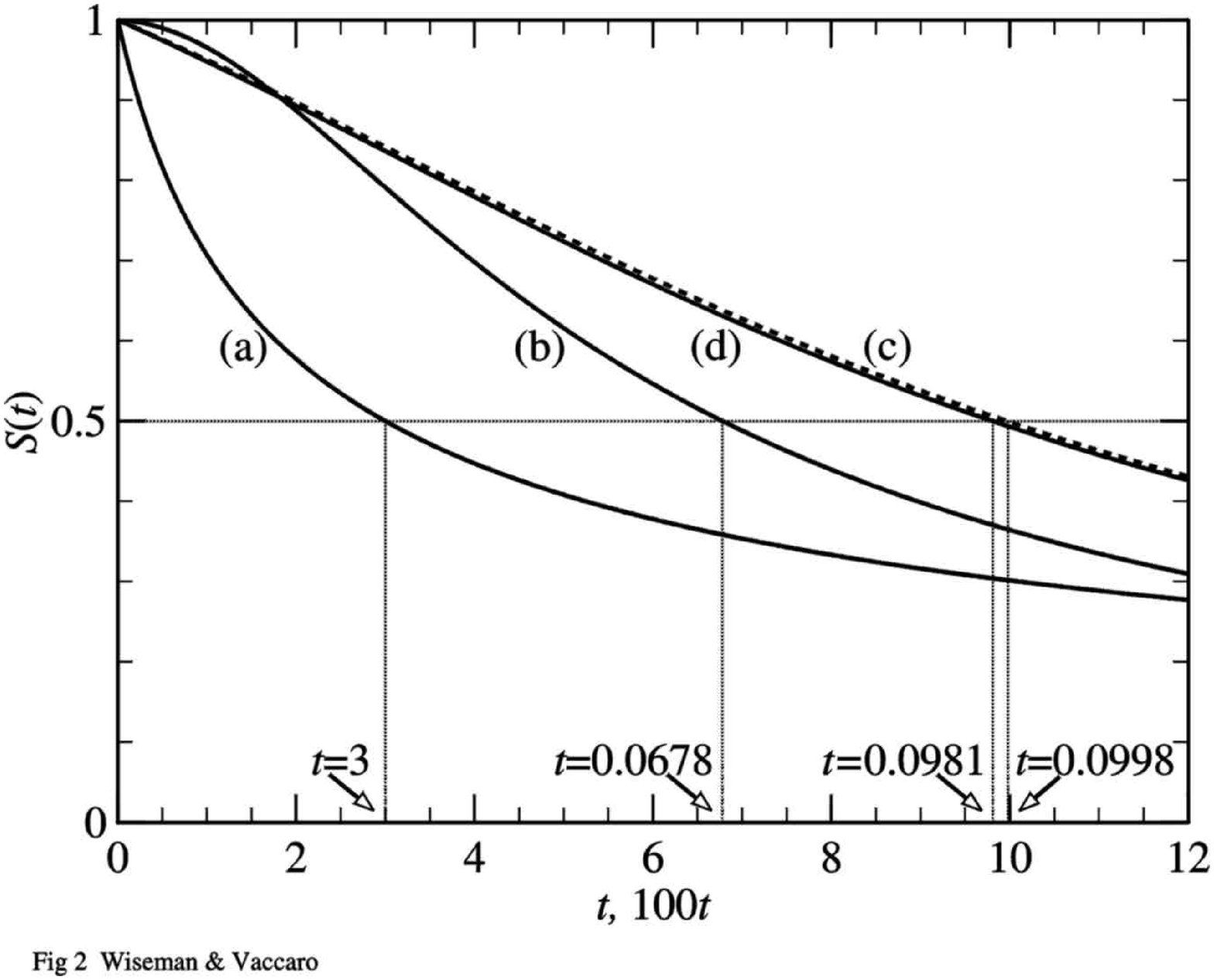,width=80mm,angle=-90,bbllx=-5mm,bblly=7mm,bburx=202mm,bbury=258mm,clip=}
\caption{\narrowtext The decay of the ensemble-averaged survival
probability in time for the four cases represented in
Fig.~\ref{fig1}. The horizontal axis measures time $t$. For case
(a) it is scaled in units of the bare lifetime of the laser mode,
and  for cases (b), (c), and (d) it is scaled in units 100 times
smaller. That is, the survival probabilities actually drop much
more quickly for the last three cases.
    \protect\label{fig2}}
\end{figure}

\begin{figure}
\psfig{figure=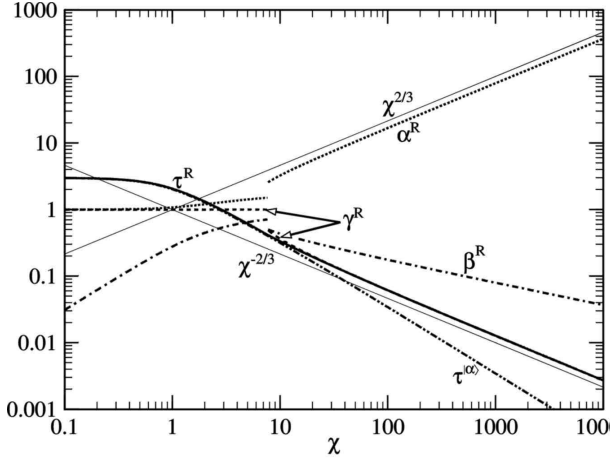,width=80mm,angle=-90,bbllx=7mm,bblly=7mm,bburx=200mm,bbury=258mm,clip=}
%bbllx=17mm,bblly=30mm,bburx=193mm,bbury=272mm,clip=}
\caption{\narrowtext The parameters for the most robust
unconstrained Gaussian ensemble $E^{\rm R}$ as a function of
$\chi$ with $\nu=0$. These parameters are the phase quadrature
variance $\alpha^{\rm R}$ (dotted line), the amplitude-quadrature
variance $\gamma^{\rm R}$  (dashed line), the covariance
$\beta^{\rm R}$ (dash-dot line) and the survival time $\tau^{\rm
R}$ for the members of this ensemble.  For comparison, we also
plot the survival time $\tau^{\ket{\alpha}}$ (dash-dot-dot line)
of a coherent state ensemble. Both survival times are in units of
the bare lifetime of the laser mode.
    \protect\label{fig3}}
\end{figure}

\begin{figure}
\psfig{figure=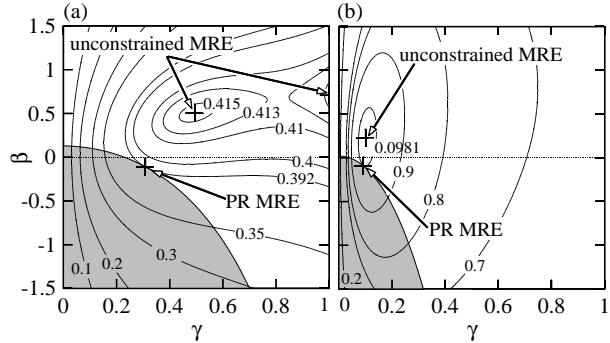,width=80mm,angle=-90,bbllx=40mm,bblly=20mm,bburx=175mm,bbury=250mm,clip=}
%bbllx=40mm,bblly=20mm,bburx=173mm,bbury=255mm,clip=}
\caption{\narrowtext Contour plots of the survival time $\tau$ as
a function of $\gamma$ and $\beta$. In (a) $\nu=0$ and $\chi=7.7$
and in (b) $\nu=0$ and $\chi=50$.  In each plot the heavy curves
represent contours of $\tau$ (in units of the bare lifetime of the
laser mode) and the shaded region represents states that are
physically realizable (PR).  Crosses mark the positions of the
maximally robust ensembles (MRE).
    \protect\label{fig4}}
\end{figure}

\begin{figure}
\psfig{figure=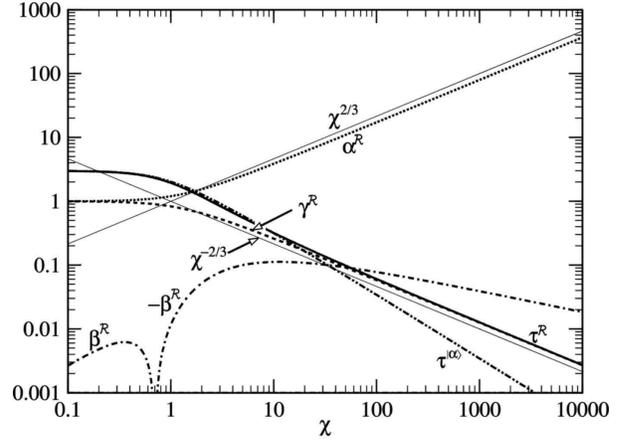,width=80mm,angle=-90,bbllx=7mm,bblly=7mm,bburx=200mm,bbury=270mm,clip=}
%angle=-90,bbllx=17mm,bblly=30mm,bburx=193mm,bbury=272mm,clip=}
\caption{\narrowtext The parameters for the ensemble $E^{\cal R}$
arising from the maximally robust unraveling ${\cal R}$ as a
function of $\chi$ with $\nu=0$. As in Fig.~\ref{fig3} we plot
$\alpha^{\cal R}$ (dotted line), $\gamma^{\cal R}$ (dashed line)
and $\pm\beta^{\cal R}$ (dash-dot lines).  We also plot the
survival time $\tau^{\cal R}$ (solid line) of this ensemble and,
for comparison, the survival time $\tau^{\ket{\alpha}}$
(dash-dot-dot line) of a coherent state ensemble. Both of these
times are in units of the bare lifetime of the laser mode.
\protect\label{fig5}}
\end{figure}

\begin{figure}
\psfig{figure=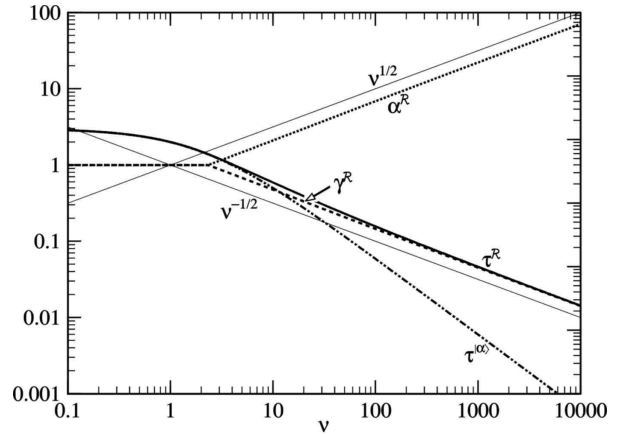,width=80mm,angle=-90,bbllx=7mm,bblly=7mm,bburx=200mm,bbury=270mm,clip=}
%bbllx=17mm,bblly=30mm,bburx=193mm,bbury=272mm,clip=}
\caption{\narrowtext The parameters for the ensemble $E^{\cal R}$
arising from the maximally robust unraveling ${\cal R}$ as a
function of $\nu$ with $\chi=0$. As in Fig.~\ref{fig3} we plot
$\alpha^{\cal R}$ (dotted line), $\gamma^{\cal R}$  (dashed line),
and the survival time $\tau^{\cal R}$ (solid line). We do not plot
$\beta^{\cal R}$ because it is identically zero. For comparison we
also plot the survival time $\tau^{\ket{\alpha}}$
(dash-dot-dotted) of a coherent state ensemble. Both of these
times are in units of the bare lifetime of the laser mode.
    \protect\label{fig6}}
\end{figure}

\begin{figure}
\psfig{figure=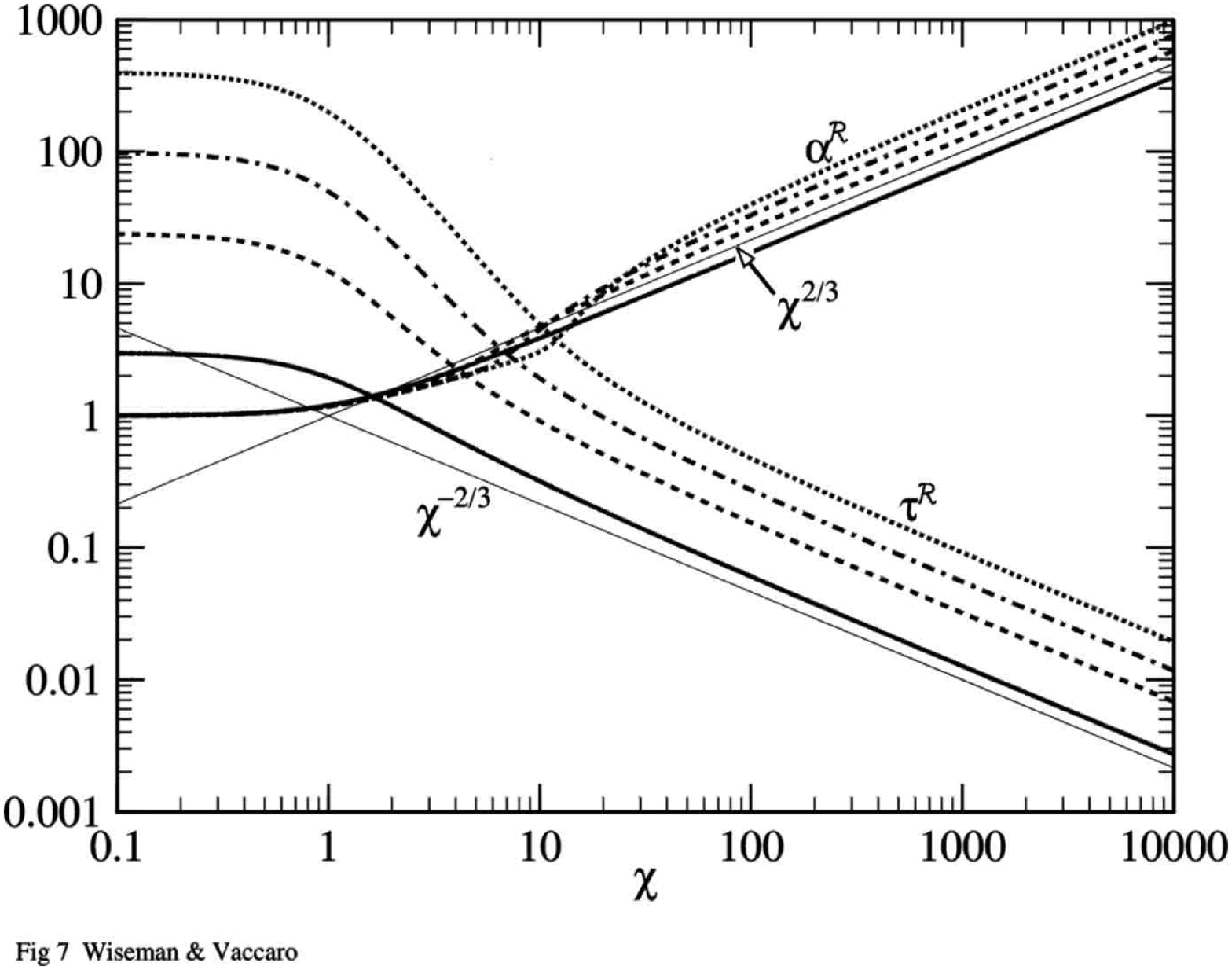,width=80mm,angle=-90,bbllx=7mm,bblly=7mm,bburx=200mm,bbury=270mm,clip=}
%bbllx=17mm,bblly=30mm,bburx=193mm,bbury=272mm,clip=}
\caption{\narrowtext The parameters for the ensemble $E^{\cal R}$
arising from the maximally robust unraveling ${\cal R}$ as a
function of $\chi$ with $\nu=0$ and for various $\Lambda$. The
rising lines are $\alpha^{\cal R}$ and the falling lines are
$\tau^{\cal R}$ (in units of the bare lifetime of the laser mode).
The values of $\Lambda$ are 0.5 (solid line), 0.2 (dashed line),
0.1 (dash-dot line), and 0.05 (dotted line).
    \protect\label{fig7}}
\end{figure}

\begin{figure}
\psfig{figure=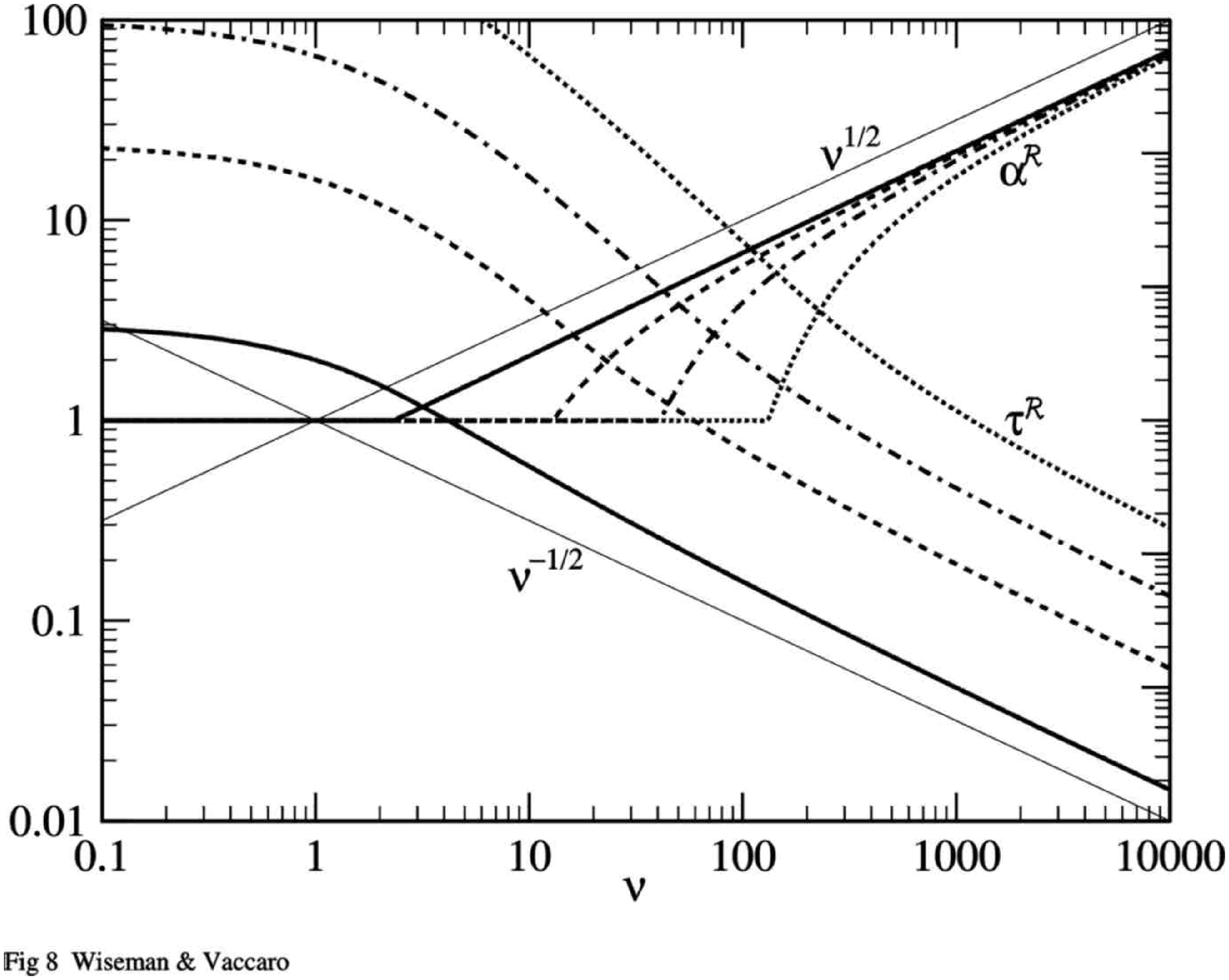,width=80mm,angle=-90,bbllx=7mm,bblly=7mm,bburx=195mm,bbury=260mm,clip=}
%bbllx=17mm,bblly=30mm,bburx=193mm,bbury=272mm,clip=}
\caption{\narrowtext The parameters for the ensemble $E^{\cal R}$
arising from the maximally robust unraveling ${\cal R}$ as a
function of $\nu$ with $\chi=0$ and for various $\Lambda$. The
rising lines are $\alpha^{\cal R}$ and the falling lines are
$\tau^{\cal R}$ (in units of the bare lifetime of the laser mode).
The values of $\Lambda$ are 0.5 (solid line), 0.2 (dashed line),
0.1 (dash-dot line), and 0.05 (dotted line).
    \protect\label{fig8}}
\end{figure}

\begin{figure}
\psfig{figure=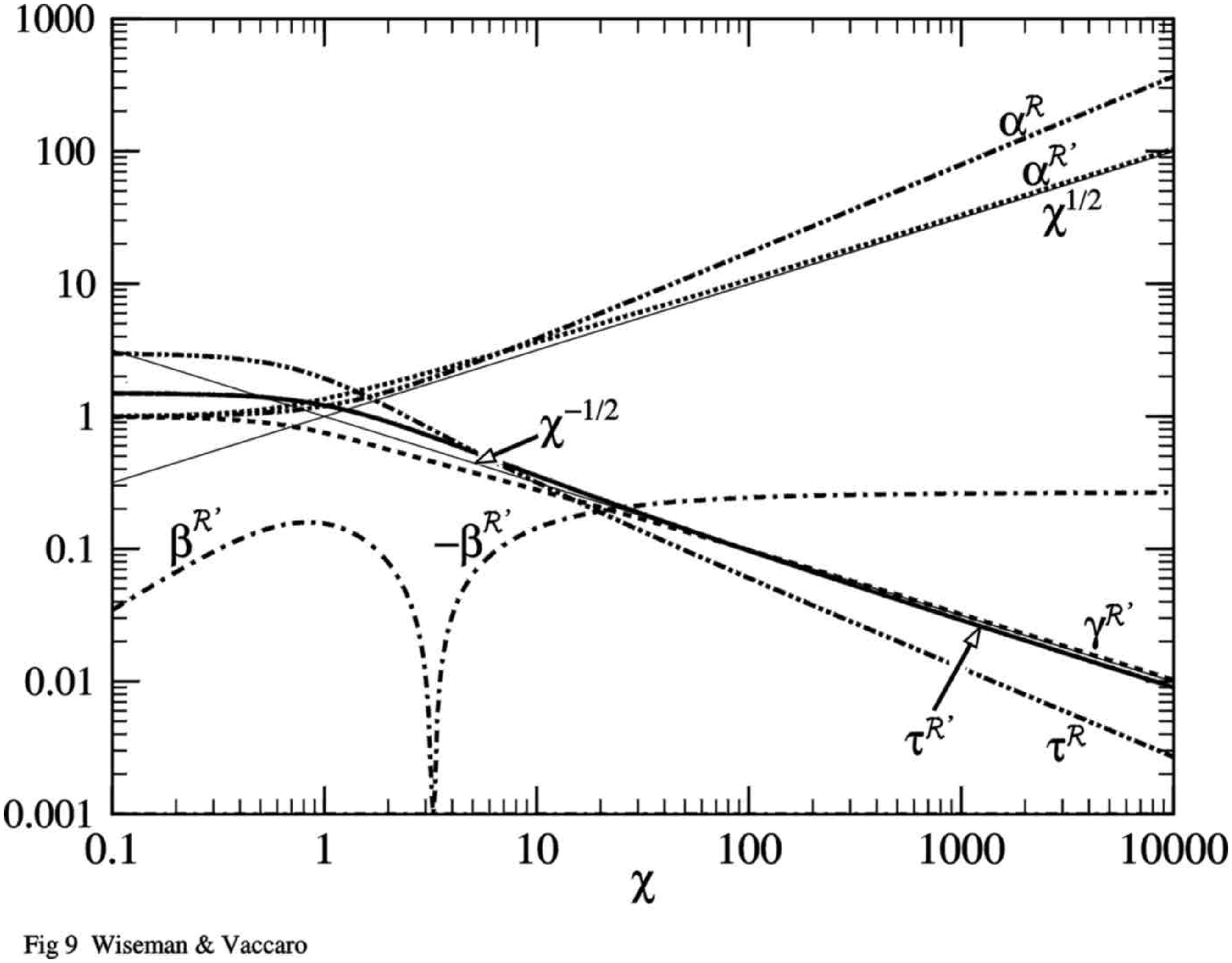,width=80mm,angle=-90,bbllx=7mm,bblly=7mm,bburx=200mm,bbury=270mm,clip=}
\caption{\narrowtext Parameters for the maximally robust ensemble
for $\nu=0$ as a function of $\chi$ as in Fig.~\ref{fig5} but
using purity as a measure of robustness.  As in previous figures
we plot $\alpha^{\cal R'}$ (dotted line), $\gamma^{\cal R'}$
(dashed line), $\beta^{\cal R'}$ (dash-dot line) and $\tau^{\cal
R'}$ (solid line). Also shown for comparison are the $\alpha^{\cal
R}$ (rising) and $\tau^{\cal R}$ (falling) curves from
Fig.~\ref{fig5} as dash-dot-dotted curves. Both times are in units
of the bare lifetime of the laser mode.
    \protect\label{fig9}}
\end{figure}

\end{multicols}


\begin{references}

\bibitem{Wis97}
H.M. Wiseman,
%``Defining the (atom) laser'',
Phys. Rev. A {\bf 56}, 2068 (1997).

\bibitem{SarScuLam74}
M.~Sargent, M.O.~Scully, and W.E.~Lamb, {\em Laser Physics}
(Addison-Wesley, Reading Mass., 1974)

\bibitem{WisVac01a}
H.M. Wiseman and J.A. Vaccaro, quant-ph/9906125 (part I).

\bibitem{Gea90}
J. Gea-Banacloche, in: {\em New Frontiers in Quantum
Electrodynamics and Quantum Optics}, A.O. Barut, ed., Plenum, New
York (1990).

\bibitem{Wis93a}
H.M. Wiseman,
%``Stochastic quantum dynamics of a continuously monitored laser''
Phys. Rev. A {\bf 47}, 5180 (1993).

\bibitem{Gea98}
J. Gea-Banacloche,
%``Emergence of Classical Radiation Fields through Decoherence \ldots''
Found. Phys. {\bf 28}, 531 (1998).

\bibitem{BarBurVac96}
S.M. Barnett, K. Burnett and J.A. Vaccaro, J. Res. Natl. Inst.
Stand. Technol. {\bf 101},593 (1996).

\bibitem{Schumacher}
B. Schumacher, Phys. Rev. A {\bf 51}, 2738 (1995).

\bibitem{WisVac98}
H.M. Wiseman and J.A. Vaccaro, Phys. Lett. A {\bf 250}, 241
(1998).

\bibitem{Zur93}
W.H. Zurek, Prog. Theor. Phys. {\bf 89}, 281 (1993).

\bibitem{ZurHabPaz93}
W.H. Zurek, S. Habib, and J.P. Paz, Phys. Rev. Lett {\bf 70},1187
(1993).

\bibitem{Gal95}
M.R. Gallis, Phys. Rev. A {\bf 53}, 655 (1996).

\bibitem{ParScu98}
Gh.-S. Paraoanu and H. Scutaru Phys. Lett. A {\bf 238}, 219
(1998).

\bibitem{DioKie00}
L. Di\'osi and C. Kiefer, Phys. Rev. Lett. {\bf 85}, 3552 (2000).

\bibitem{DalDziZur01}
D. Dalvit, J. Dziarmaga, and W.H. Zurek, Phys. Rev. Lett. {\bf
86}, 373 (2001).

\bibitem{WisBra00}
H.M. Wiseman and Z. Brady, Phys. Rev. A {\bf 62}, 023805 (2000).

\bibitem{Car93b}
H.J. Carmichael, {\em An Open Systems Approach to Quantum Optics}
(Springer-Verlag, Berlin, 1993).

\bibitem{Lin76}
G. Lindblad, Commun. math. Phys. {\bf 48}, 199 (1976).

\bibitem{fidelity}
In Ref.~\cite{BarBurVac96} we used the same overlap as in
\erf{overlap} as a measure of robustness, which, following
Ref.~\cite{Schumacher}, we referred to as fidelity.  However, we
feel that the term {\em survival probability} is more appropriate
since it embodies the notion of the state $P_n^{\cal U}$ to
survive over time $t$ whereas {\em fidelity} embodies the faithful
reproduction of a state.

\bibitem{Hil84}
M. Hillery, R.F. O'Connell, M.O. Scully, and E.P. Wigner, Phys.
Rep. {\bf 106}, 121 (1984).

\bibitem{GisPer92}
N. Gisin and I. Percival, Phys. Lett. A {\bf 167}, 315 (1992);
{\em ibid.}, J. Phys. A {\bf 25}, 5677 (1992).

\bibitem{Dio88}%QSD
L. Di\'osi, Phys. Lett. A {\bf 132}, 233 (1988).

\bibitem{Ang97}
J. Anglin, Phys. Rev. Lett. {\bf 79}, 6 (1997).

\bibitem{Bra00}
Z. Brady, B.Sc. Honours thesis (Griffith University, 2000).

\end{references}
\end{document}